\documentclass[12pt]{iopart}

\usepackage{iopams}
\usepackage{amssymb}
\usepackage{amsfonts}
\usepackage{bm}
\usepackage{epsfig}
\usepackage{epsf}
\usepackage[]{graphicx}

\usepackage{color}

\usepackage{eurosym}

\usepackage{cite}
\usepackage{xspace}

\usepackage{ulem}

\renewcommand{\d}{{\rm d}}
\renewcommand{\i}{{\rm i}}
\newcommand {\ee}{{\rm e}}

\newcommand{\E}  {{\varepsilon}}
\newcommand{\om}  {{\omega}}
\newcommand{\Om}  {{\Omega}}

\newcommand{\Ld}  {L_{\rm d}}
\newcommand{\Lb}  {L_{\rm b}}
\newcommand{\La}  {L_{\rm a}}
\newcommand{\Ldm} {L_{\rm dm}}

\newcommand{\Nd}   {N_{\rm d}}
\newcommand{\Ndm}  {N_{\rm dm}}
\newcommand{\Neff} {{N_{\mathrm{eff}}}}
\newcommand{\Nu}   {{N_{\rm u}}}

\newcommand{\coh}  {{\rm coh}}

\newcommand{\lamu} {{\lambda_{\rm u}}}

\newcommand{\Eph} {{E_{\rm ph}}}

\newcommand{\calN} {{\cal N}}
\newcommand{\calA} {{\cal A}}

\newcommand{\aTF}  {a_{\rm TF}}

\newcommand{\dUmax}  {U_{\max}^{\prime}}

\begin{document}
%\jl{2}

% Title, authors

%%%%%%%%%%%%%%%%%%%%%%%%%%%%%%%%%%%%%%%%%%%%%%%%%%%%%%%%%%%%%%%%
\title[Crystal-Based Light Sources]
{Crystal-based intensive gamma-ray light sources}

\author{Andrei V. Korol and Andrey V. Solov'yov$^*$\footnote{On leave 
from A.F. Ioffe Physical-Technical Institute, St. Petersburg, Russia}}

\address{
MBN Research Center, Altenh\"{o}ferallee 3, 60438 
Frankfurt am Main, Germany}

%\ead{korol@mbnexplorer.com}
\ead{solovyov@mbnresearch.com}

%%%%%%%%%%%%%%%%%%%%%%%%%%%%%%%%%%%%%%%%%%%%%%%%%%%%%%%%%%%%%%%%%
\begin{abstract}
We discuss design and practical realization of novel gamma-ray Crystal-based Light Sources (CLS) 
that can be constructed through exposure of oriented crystals (linear, bent, periodically bent) 
to beams of ultrarelativistic charged particles. 
In an exemplary case study, we estimate brilliance of radiation emitted in a Crystalline Undulator 
(CU) LS by available positron beams.
Intensity of CU radiation in the photon energy range $10^0-10^1$ MeV, which is inaccessible to 
conventional synchrotrons, undulators and XFELs, greatly exceeds that of 
laser-Compton scattering LSs and can be higher than predicted in the Gamma Factory proposal to CERN.
Brilliance of CU-LSs can be boosted by up to 8 orders of magnitude through the process of superradiance 
by a pre-bunched beam. 
Construction of novel CLSs is a challenging task which constitutes a highly interdisciplinary field 
entangling a broad range of correlated activities.
CLSs provide a low-cost alternative to conventional LSs and have enormous number of applications.
\end{abstract}

%%%%%%%%%%%%%%%%%%%%%%%%%%%%%%%%%%%%%%%%%%%%%%%%%%%%%%%%%%%%%%
\section{Introduction \label{Introduction}}

The development of light sources (LS) for wavelengths $\lambda$
well below 1 angstrom 
(corresponding photon energies $\Eph \gg 10$ keV) 
is a challenging goal of modern physics.
Sub-angstrom wavelength powerful spontaneous and, especially, coherent radiation will
have many applications in the basic sciences, technology and medicine.
They may have a revolutionary impact on nuclear and solid-state physics,
as well as on the life sciences.
At present, several X-ray Free-Electron-Laser (XFEL) 
sources are operating (European XFEL, FERMI, LCLS, SACLA, PAL-XFEL) or planned (SwissFEL) 
for X-rays down to $\lambda \sim 1$ \AA{}  
 \cite{Doerr-EtAl_NatureMethods_v15_p33_2018,Seddon-EtAl_RepProgPhys_v80_115901_720_2017,
 SwissFEL_ApplScie_v7_720_2017,Bostedt-EtAl_RMP_v88_015007_2016,
 Emma-EtAl_NaturePhotonics_v4_015006_2010, McNeilThompson_NaturePhotonics_v4_p814_2010}.
However, no laser system has yet been commissioned for lower wavelengths
due to the limitations of permanent magnet and accelerator technologies.
Modern synchrotron facilities, such as APS, SPring-8, PETRA III, ESRF
\cite{YabashiTanaka_NaturePhotonics_v11_p12_2017,Ayvazyan_EtAl_EPJD_v20_p149_2002}, 
 provide radiation of shorter wavelengths but of much less intensity which 
 falls off very rapidly as $\lambda$ decreases.

Therefore, to create a powerful LS in the range well below 1 \AA{}, 
i.e. in the hard X and gamma ray band, 
one has consider new approaches and technologies.

%%%%%%%%%%%%%%%%%%%%%%%%%%%%%%%%%%%%%%%%%%%%%%%%%%%%%%%%%%%%%%%%%%%%%%%%%%%%%%%%

In this article we discuss possibilities and perspectives for designing and 
practical realization of novel gamma-ray Crystal-based LSs (CLS) operating at 
photon energies $\Eph\gtrsim10^2$ keV and above that can be constructed through exposure of 
oriented crystals 
(linear, bent and periodically bent crystals) 
%(linear Crystals -- LC, Bent Crystals -- BC, Periodically Bent Crystals -- PBC) 
to beams of ultrarelativistic charged particles. 
CLSs include Channeling Radiation (ChR) emitters, crystalline 
synchrotron radiation emitters, 
%Synchrotron Radiation (SR) emitters, 
crystalline Bremsstrahlung %(BrS) 
radiation emitters,
Crystalline Undulators (CU) and stacks of CUs. 
This interdisciplinary research field combines theory, computational modeling, beam manipulation,
design, manufacture and experimental verification of high-quality crystalline samples and 
subsequent characterization of their emitted radiation as novel LSs.  
In an exemplary case study, we estimate the characteristics
(brilliance, intensity) of radiation emitted in CU-LS by positron beams 
available at present.
It is demonstrated that peak brilliance of the CU Radiation (CUR) 
at $\Eph=10^{-1}-10^2$ MeV is comparable to   
or even higher than that achievable in conventional synchrotrons but for much  
lower photon energies.  
Intensity of radiation from CU-LSs greatly exceeds that available in the 
laser-Compton scattering LSs and can be made higher than predicted in the 
Gamma Factory proposal to CERN \cite{GammaFactory_CERN-Courier,Krasny:2018xxv}.
The brilliance can be boosted by orders of magnitude 
through the process of superradiance by a pre-bunched beam. 
We show that brilliance of superradiant CUR can be comparable with the values achievable 
at the current XFEL facilities which operate in much lower photon energy range. 

%%%%%%%%%%%%%%%%%%%%%%%%%%%%%%%%%%%%%%%%%%%%%%%%%%%%%%%%%%%%%%%%%%%%%%%%%%%%%%

CLSs can generate radiation in the photon energy range 
where the technologies based on the charged particles motion in the fields of permanent magnets 
become inefficient or incapable. 
The limitations of conventional LS is overcome by exploiting very strong crystalline fields
that can be as high $\sim10^{10}$ V/cm, which is equivalent
to a magnetic field of 3000 Tesla whilst modern superconducting magnets provide
1-10 Tesla \cite{ParticleDataGroup2014}. 
The orientation of a crystal along the beam enhances significantly 
the strength of the particles interaction with the crystal due to strongly correlated 
scattering from lattice atoms. 
This allows for the guided motion of particles through crystals of different geometry and for 
the enhancement of radiation.

\begin{figure} [h]
\centering
\includegraphics[width=15cm,clip]{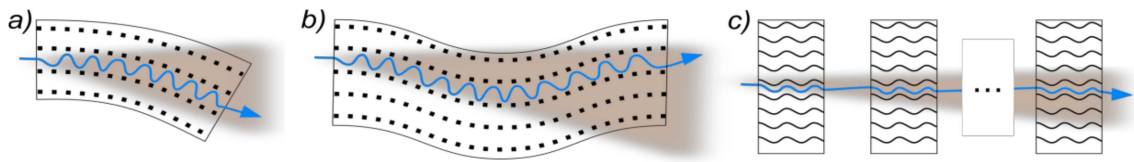}%{CLS_v02.eps}
\caption{
Selected examples of the novel CLSs: 
%(a) BC, (b) PBC, (c) a stack of PBCs.
(a) bent crystal, (b) periodically bent crystal, (c) a stack of periodically bent crystals.
Black circles and lines mark atoms of crystallographic
planes, wavy curves show trajectories of the channeling particles, 
shadowed areas refer to the emitted radiation.}
\label{Figure01.fig}%{CLS.fig}
\end{figure}

Examples of CLSs are shown in Figure \ref{Figure01.fig}. 
%The SR
The synchrotron radiation is emitted by ultra-relativistic projectiles propagating in 
the channeling regime through a % BC
bent crystal, panel a). 
A CU, panel b), contains a %PBC 
periodically bent crystal and a beam of channeling particles which emit CUR following the periodicity of the 
%crystal 
bending  \cite{KSG1998,KSG_review_1999,KSG_review_2004}. 
A CU-based LS can generate photons of $\Eph=10^2$ keV - $10^1$ GeV 
 range (corresponding to $\lambda$ from 0.1 to $10^{-6}$ \AA). 
Under certain conditions, CU can become a source of the coherent light within the 
range $\lambda = 10^{-2}-10^{-1}$ \AA{}
\cite{KSG_review_2004,Patent,ChannelingBook2014}.
An LS based on a stack of CUs is shown in panel c) \cite{Sushko_AK_AS_SPB_SASP_2015}.

%%%%%%%%%%%%%%%%%% 
% %%%%%%%%%%%%%%%%%%%%%%%%%%%%%%%%%%%%%

Practical realization of CLSs often relies on the channeling effect. 
The basic \textit{phenomenon of channeling} is in a large distance which a projectile particle penetrates 
moving along a crystallographic plane or axis and experiencing collective action of the electrostatic 
fields of the lattice atoms \cite{Lindhard}. 
A typical distance covered by a particle before it leaves the channeling mode
due to uncorrelated collisions is called the \textit{dechanneling length}, $\Ld$.
It depends on the type of a crystal and its orientation, on the type of channeling motion, planar or axial, 
and on the projectile energy and charge. 
In the planar regime, positrons channel in between two adjacent planes whereas electrons propagate in 
the vicinity of a plane thus experiencing more frequent collisions. 
As a result, $\Ld$ for electrons is much less than for positrons.
To ensure enhancement of the emitted radiation due to the dechanneling effect, the crystal length $L$ must 
be chosen as $L \sim\Ld$ \cite{KSG1998,KSG_review_1999,KSG_review_2004}. 

\begin{figure} [h]
\centering
\includegraphics[width=10cm,clip]{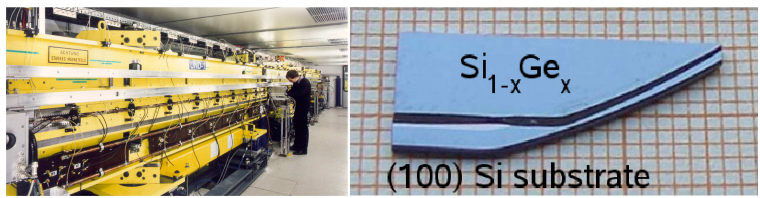}%{DESY-vs-CU.eps}
\caption{
\textit{Left:} 
Magnetic undulator for the European XFEL.
\textit{Right:} A Si$_{1-x}$Ge$_x$ superlattice CU (the upper layer) build atop the 
silicon substrate (the lower layer) with the face normal to the $\langle 100 \rangle$ 
crystallographic direction.   
In the superlattice, the (110) planes are bent periodically.
The picture courtesy of
J.L. Hansen, A. Nylandsted and U. Uggerh\o{}j (University of Aarhus). 
The whole figure is adapted from Ref. \cite{MBNExplorer_Book}.
}
\label{Figure02.fig}%{DESY-vs-CU.fig}
\end{figure}

%%%%%%%%%%%%%%%%%%% 
The motion of a projectile and the radiation emission in 
% BC and PBC
bent and periodically bent crystals 
are similar to those in magnet-based synchrotrons and undulators. 
The main difference is that in the latter the particles and photons move in vacuum whereas in crystals they 
propagate in medium, thus leading to a number of limitations for the crystal length, bending curvature,  
and beam energy.
However, the crystalline fields are so strong that they 
steer ultra-relativistic particles more effectively than the most advanced magnets. 
Strong fields bring bending radius in bent crystals
%BC 
down to the cm range and bending period $\lamu$ in periodically bent crystals
%PBC 
to the hundred or even ten microns range. 
These values are orders of magnitude smaller than those achievable with magnets
\cite{Emma-EtAl_NaturePhotonics_v4_015006_2010}.
As a result, the radiators can be miniaturized thus lowering dramatically the cost of 
CLSs as compared to that of conventional LSs. 
Figure \ref{Figure02.fig} matches the magnetic undulator for the European XFEL with the CU manufactured in 
University of Aarhus and used  in recent experiments \cite{Bandiera_etal:PRL_v115_025504_2015}.

Modern accelerator facilities make available intensive electron and
positron beams of high energies, from the sub-GeV up to hundreds of GeV. 
These energies combined with large bending curvature achievable in crystals provide a possibility to consider 
novel CLSs of the synchrotron type (continuous spectrum radiation) and of the undulator type 
(monochromatic radiation) of the energy range up to tens of GeV.
Manufacture of high quality 
%BC and PBC 
bent and periodically bent crystals
is at the edge of current technologies.

%%%%%%%%%%%%%%%%%% 
% %%%%%%%%%%%%%%%%%%%%%%%%%%%%%%%%%%%%%

%%%%%%%%%%%%%%%%%%%%%%%%%%%%%%%%%%%%%%%%%%
\section{Exemplary crystal-based LSs \label{CLS}}

A number of theoretical and experimental studies of the channeling phenomenon in oriented 
%LCs 
linear crystals have been carried out  (see, e.g., a review \cite{UggerhojRPM}).
A channeling particle emits intensive ChR, which was predicted theoretically \cite{ChRad:Kumakhov1976} and 
shortly after observed experimentally \cite{ChRad-Exp:TerhunePantell_APL_v30_p265_1977}.
Since then there has been extensive theoretical and experimental investigation of ChR.
The energy of emitted photons $E_{\rm ph}$ scales with the beam energy 
as $\E^{3/2}$ and thus can be varied by changing the latter.
For example, by propagating electrons of moderate energies, $\E=10-40$ MeV, through 
%LC 
a linear crystal it is possible to generate ChR with photon energy $\Eph =10-80$ keV  
\cite{Brau_EtAl-SynchrRadNews_v25_p20_2012,Wagner_EtAl-NIMB_v266_p327_2008}.
This range can be achieved in magnetic undulators but with much higher beam energy.
High-quality electron beams of (tunable) energies within the tens of MeV range 
are available at many facilities.
Hence, it has become possible to consider ChR from 
%LC
linear crystals
as a new powerful LS in the X-ray range \cite{Brau_EtAl-SynchrRadNews_v25_p20_2012}.

In the gamma-range, ChR can be emitted by higher energy $\E\gtrsim 10^2$ MeV beams. 
However, modern accelerator facilities operate at a fixed value of $\E$
(or, at several fixed values) \cite{ParticleDataGroup2018,SuperKEKB-Collider_ArXiv_1809.01958v1_2018,
CEPC-Design-Report_arXiv1809-00285_2018}. 
This narrows the options for tuning the ChR parameters, in particular, the wavelength.
Hence, the corresponding CLS lack the tunability option.
From this viewpoint, the use of 
%BCs and, especially, PBCs
bent and, especially, periodically bent crystals
can become an alternative as they provide tunable emission in the  hard X- and gamma-ray range.

Strong crystalline fields give rise to channeling in a bent crystal. 
%BC.
Since its prediction \cite{Tsyganov1976} and experimental support \cite{ElishevEtAl:PLB_v88_p387_1979}, the 
idea to deflect high-energy beams of charged particles by means of bent crystals
%BCs 
has attracted a lot of attention
\cite{BiryukovChesnokovKotovBook,UggerhojRPM}.
The experiments have been carried with ultra-relativistic protons,
ions, positrons, electrons, $\pi^{-}$-mesons 
\cite{Scandale_etal:PL_B719_p70_2013,ScandaleEtAl:PRSTAB_v11_063501_2008,%
ScandaleEtAl:PRA_v79_012903_2009,Bandiera_etal:PRL_v115_025504_2015,FlillerEtAl:PRSTAB_v9_013501_2006}.
Steering of highly energetic electrons and positrons in 
%BC 
bent crystals with small bending radius 
$R$ gives rise to intensive 
%SR
synchrotron radiation with $\Eph\gtrsim 10^0$ MeV. 
The parameters of radiation can be tuned by varying $R$ within the range $10^0-10^2$ cm 
\cite{Mazzolari_etal:PRL_v112_135503_2014,ShenEtAl_NIMB-v424-p26-2018}.

Even more tunable is a CU-LS.
In this system CUR and ChR are emitted in distinctly different photon energy ranges
so that CUR is not affected by ChR. 
The intensity, photon energy and line-width of CUR can be varied and tuned 
by changing $\E$, bending amplitude $a$ and period $\lamu$, 
type of crystal, its length and detector aperture \cite{ChannelingBook2014}. 

%%%%%%%%%%%%%%%%%
Since introducing the concept of CU, major theoretical studies have been devoted to the large-amplitude 
large-period bending $\lamu \gg a>d$ \cite{KSG1998,KSG_review_1999,KSG_review_2004}.
In this regime, a projectile follows the shape of periodically bent planes.
CUR is emitted at the frequencies $\om_{\rm u}$ well below those of ChR, $\om_{\rm ch}$.
By varying $a$,  $\lamu$, $\E$ and the crystal length one can tune the CUR peaks positions and intensities.
Small-amplitude small-period regime, which implies $a \ll d$ and $\lamu$ less than period of channeling oscillations
\cite{Kostyuk_PRL_2013,Wistisen_etal:PRL_v112_254801_2014,Wistisen_etal:EPJD_2017,KorolBezchastnovSushkoSolovyov2016,
UggerhojWistisen:NIMB_v355_p35_2015}.
This scheme  allows emission of photons of the higher energies, $\om_{\rm u}>\om_{\rm ch}$,
makes feasible construction of a CLS which radiates in the GeV photon energy range
\cite{Bezchastnov_AK_AS:JPB_v47_195401_2014}.

Initially, the CU feasibility was justified for positrons \cite{KSG1998,KSG_review_1999}.
Positrons channel over larger distances passing larger number of
bending periods and, thus, increasing the CUR intensity.
Experiments carried out so far to detect CUR from positrons have not been successful due insufficient 
quality of periodic bending, large beam divergence and high level of the background bremsstrahlung 
radiation \cite{Baranov_etal:NIMB_v252_p32_2006,Backe_etal:NuoCimC_v34_p175_2011}.
The feasibility of CU for electrons was also proven \cite{PRL2007} but it was indicated that to obtain 
better CUR signal high-energy (GeV and above) electron beams are preferable.
The CUR signal was detected in the experiments with electron beam of much lower energies 
at the Mainz Microtron \cite{Backe_EtAl_2011,Backe_EtAl_2013}.
% The radiation excess due to CUR was detected although it was not as intense as expected
% due to insufficient quality of the crystalline lattice.
\textcolor{black}{
The radiation excess due to CUR was detected although it was not as intense as expected. 
In part, this discrepancy can be attributed to insufficient quality of the crystalline lattice although this issue has to be
investigated in more detail.
}
Also the beam energy used was low (sub-GeV range) and as a consequence photon 
energies, as well as the choice of particles (electrons) were not optimal.

%%%%%%%%%%%%%%%%%%%%%%%%%%%%%%%%%%%%%%%%%%%%%%%%
\section{Practical realization of CU 
 \label{Feasibility-Methods}} 

In the current paper a case study of a tunable \textit{CU-based} LS is presented. 
Therefore below we focus on the methods which allow one to produce undulating crystal samples.  
 
The feasibility of the CU concept was verified theoretically in Refs.
\cite{KSG1998,KSG_review_1999,KSG_review_2004,EnLoss00} where essential conditions and 
limitations  which must be met were formulated.
These papers boosted theoretical and experimental investigation of the CU and CUR phenomena 
worldwide, so that nowadays these topics represent a new and very rich field of research.

Theoretical and experimental studies of the CU and CUR phenomena has ascertained the importance of 
the high quality of the undulator material needed to achieve strong effects in the emission spectra.  
Up to now, several methods to create 
%PBC 
periodically bent crystalline 
structures have been proposed and/or realized.

\begin{figure} [ht]
\centering
\includegraphics[width=10cm,clip]{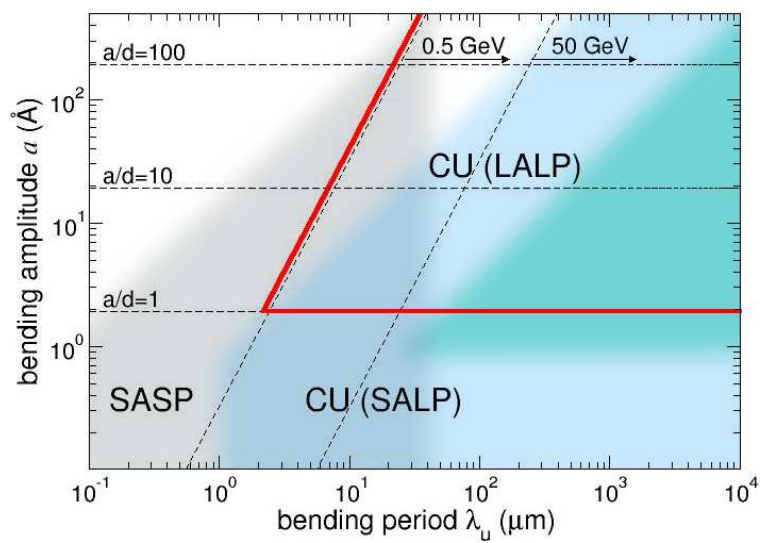}%{Figure02_v06.eps}
\caption{
Shadowing indicates the ranges accessible by 
means of modern technologies: superlattices (grey), surface deformations (green), 
acoustic waves (blue). 
Sloping dashed lines indicate the boundaries 
of the stable channeling for $\E = 0.5$ and 50 GeV projectiles.
For each energy, the periodic bending corresponding to the CU regime (characterized by Large Period, (LP),
$\lamu \gg a$) lies to the right from the line.
The horizontal line $a/d = 1$ ($d$ is the interplanar spacing) separates the Large-Amplitude (LA)
and Small-Amplitude (SA) bending.
The boundaries of the most favourable CU regime, LALP, are marked by thick red lines.
%See also explanation in the text.
}
\label{Figure03.fig}%{Figure02_v06.fig}
\end{figure}

Figure \ref{Figure03.fig} provides schematic illustration of the ranges of $a$ and
$\lamu$ within which the emission of intensive CUR is feasible. 
Shadowed areas mark the ranges currently achievable by different technologies.

Several approaches have been applied to produce \textit{static} bending.
The greenish area marks the area achievable by means of technologies 
based on surface deformations.
These include mechanical scratching \cite{Bellucci_etal:PRL_v90_034801_2013}, 
laser ablation technique \cite{Balling_etal:NIMB_v267_p2952_2009},
grooving method \cite{Guidi_etal:NIMB_v234_p40_2005,Bagli_etal:EPJC_v74_p3114_2014},
tensile/compressive strips deposition
\cite{Guidi_etal:NIMB_v234_p40_2005,Guidi_etal:APL_v90_114107_2007,Guidi-EtAl_ThinSolFilms_v520_p1074_2011},
ion implantation \cite{Bellucci_EtAl-APL_v107_064102_2015}.
The most recent techniques proposed is based on sandblasting one of the major sides of a crystal to produce an 
amorphized layer capable of keeping the sample bent \cite{Camattari-EtAl:JApplCryst_v50_p145_2017}.
Another technique, which is under consideration in for manufacturing periodically bent silicon and germanium crystals, 
is pulsed laser melting processing that produces localized and high-quality stressing alloys on the crystal surface. 
This technology is used in semiconductor processing to introduce
foreign atoms in crystalline lattices \cite{Cristiano-EtAl:MatScieSemicondProc_v42_p188_2018}.
Currently, by means of the surface deformation methods the 
% PBC
periodically bent crystals with large period, 
$\lamu\gtrsim10^2$ microns, can be produced. 

To decrease the period $\lamu$ one can rely on production of graded composition strained layers in an epitaxially 
grown Si$_{1-x}$Ge$_x$ superlattice \cite{BogaczKetterson1986,MikkelsenUggerhoj:NIMB_v160_p435_2000}. 
Both silicon and germanium crystals have the diamond structure with close 
lattice constants. 
Replacement of a fraction of Si atoms with Ge atoms leads to bending 
crystalline directions. 
By means of this method sets of 
%PBC 
periodically bent crystals
have been produced and used in channeling experiments 
 \cite{Backe_EtAl-NIMB_v309_p37_2013}. 
A similar effect can be achieved by graded doping during 
synthesis to produce diamond superlattice \cite{ThuNhiTranThi_JApplCryst_v50_p561_2017}. 
Both boron and nitrogen are soluble in diamond, however, higher concentrations of boron 
can be achieved before extended defects appear 
\cite{ThuNhiTranThi_JApplCryst_v50_p561_2017,Guzman_etal:DiamondRelMat_v16_p809_2007}.
The advantage of a diamond crystal is radiation hardness allowing it
to maintain the lattice integrity in the environment of very intensive 
beams \cite{UggerhojRPM}.
The grey area in \ref{Figure03.fig} marks the ranges of parameters achievable by means of crystal growing. 

The bluish area indicates the range of parameters achievable by means of another method,
realization of which is although still pending, based on propagation of a transverse 
acoustic wave %(AW) 
in a crystal \cite{ChannelingBook2014}.

%%%%%%%%%%%%%%%%%% 
%%%%%%%%%%%%%%%%%%%%%%%%%%%%%%%%%%%%%%

In a Crystalline Undulator (CU), a projectile's trajectory follows the profile of periodic bending.
This is possible when the electrostatic crystalline field exceeds the centrifugal force acting on the projectile. 
This condition, which entangles bending amplitude and period, the projectile's energy and the crystal field 
strength, implies that the bending parameter $C$ is less than one, see Eq. (\ref{Optimal_Length_CU:eq.04}).
Two sloping dashed lines in Fig. \ref{Figure03.fig} show the dependences $a=a(\lamu)$ 
corresponding to the extreme value $C=1$ for $\E=0.5$ and 50 GeV projectiles. 
For each energy, the CU is feasible in the domain lying to the right from the line.
In this domain, periodic bending is characterized by a Large Period (LP), which implies (i) $\lamu\gg a$, and
(ii) $\lamu$ greatly exceeds the period of channeling oscillations. 
The horizontal line $a/d=1$ ($d$ stands for the interplanar distance) divides the CU domain into
two parts: the Large-Amplitude (LA), $a > d$, and the Small-Amplitude (SA), $a<d$, regions.
Larger amplitudes are more favourable from the viewpoint of achieving higher intensities of CUR.
The red lines delineate the domain where the LALP periodic bending can be considered.

Another regime of periodic bending, Small-Amplitude Short-Period (SASP), can be realized in the domain
$a < d$ and $\lamu < 1$ micron (these values of $\lamu$ are much smaller that channeling periods
of projectiles with $\E \gtrsim 1$ GeV).
In the SASP regime, in contrast to the channeling in a CU, channeling particles do not follow the short-period bent 
planes but experience regular jitter-type modulations of their trajectories 
which lead to the emission of high-energy radiation. 

%%%%%%%%%%%%%%%%%%%%%%%%%%%%%%%%%%%%%  AW %%%%%%%%%%%%%%%%%%%%%%%%%%%%%%%%%%%%%%%%%%%%%%%%
As mentioned,  \textit{dynamic} bending can be achieved by propagating a transverse 
%AW 
acoustic wave along a particular crystallographic direction
\cite{KaplinPlotnikovVorobev1980,BaryshevskyDubovskayaGrubich1980,KSG1998,KSG_review_1999,IkeziLinLiuOhkawa1984,
MkrtchyanEtal1988,Dedkov1994}.
This can be achieved, for example, by placing a piezo sample atop the crystal and generating radio frequencies to excite the oscillations.
The advantage of this method is in its flexibility: the bending amplitude and period can be changed by varying the 
% AW
wave intensity and frequency \cite{KSG1998,KSG_review_1999}.
Although the applicability of this method has not yet been checked experimentally,
we note that a number of experiments has been carried out on the stimulation of ChR by
%AW
acoustic waves excited in piezoelectric crystals \cite{Wagner_EtAl_2011}.

% %%%%%%%%%%%%%%%%%% 
\begin{figure} [h]
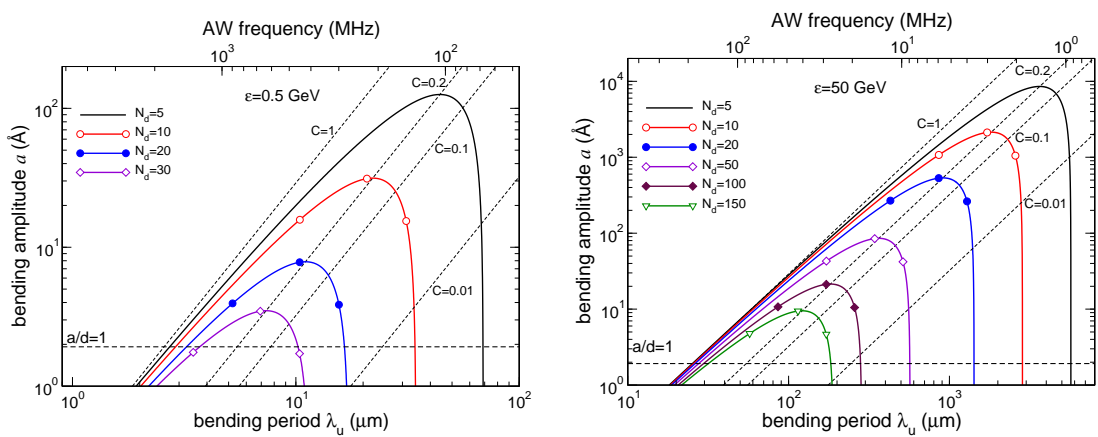

\centering
\includegraphics[width = 7cm,clip]{Figure04a.eps}%{p-0_5GeV_a_vs_lamu_v02.eps}
\hspace*{0.5cm}%
\includegraphics[width = 7cm,clip]{Figure04b.eps}%{p-50GeV_a_vs_lamu_v02.eps}
\caption{
Ranges of 
% AW 
acoustic wave frequency $\nu$ ({upper horizontal axis}), of bending period equal to the 
%AW 
wave wavelength $\lamu=\lambda_{\rm AW}$ ({lower horizontal axis}) and of amplitude $a$ ({vertical axis}) 
to be probed to construct a silicon(110)-based CU in the LALP regime.
The data refer to $\E=0.5$ GeV (left panel) and $\E=50$ GeV positrons.}
\label{Figure04.eps}
\end{figure}

Figure \ref{Figure04.eps} allows one to estimate the 
%AW 
acoustic wave frequencies $\nu$ needed
to achieve the LALP periodic bending of the silicon(110) planes.
The diagonal dashed lines correspond to the dependences $a=a(\lamu)$
obtained for several values (as indicated) of the bending parameter $C$.
The CU cannot be realized in the domain lying to the left from the line $C=1$.
The solid curves present the dependences  $a=a(\lamu)$ calculated for the fixed values (indicated in 
the legends) of the number of undulator periods $N_{\rm d}=\Ld(C)/\lamu$ within the dechanneling 
length $\Ld(C)$, Eq. (\ref{Optimal_Length_CU:eq.05}).
It is seen that the values $\lamu \sim 1\dots 10^3$ microns correspond to the frequencies
$\nu=v_{\rm s}/\lamu \sim  1\dots 10^3$\,MHz, which are achievable 
experimentally ($v_{\rm s}=4.67\times 10^5$ cm/s is the speed of sound)
\cite{Wagner_EtAl_2011,Tzianaki-EtAl:OptExpress_v23_p17191_2015,Bakarezos-EtAl:Ultrasonics_v86_p14_2018}. 

%%%%%%%%%%%%%%%%%%%%%%%%%%%%%%%%%%%%%%%%%%%%%%%%%%%%%%%%%%%%%%%%%%%%%%%%%%%%%%%%%%%%%%%%%
\section{CU-LS versus state-of-the-art LS
 \label{Results}} 

In this section, we present quantitative estimates for the CUR {\it brilliance} using the 
parameters of high-energy positron beams either available at present or planned to be 
commissioned in near future (see Table \ref{ep-beams-2018.Table} in \ref{Beams2018}).
We demonstrate that 
%by means of CU-LS 
by means of CU-LS\textcolor{black}{, which operates in the LALP regime,}  
one can achieve much higher photon yield as compared to the values 
achievable in modern LS facilities operating in the gamma-ray range,  $E_{\rm ph} \gtrsim 10^2$ keV.
 
The relevant modern facilities are synchrotrons and undulators based on the action of magnetic 
field.\footnote{Fore the sake of comparison we also match our data to the brilliance available 
at the XFEL facilities for much lower energy of the emitted radiation.} 
Another type of modern LS, which does not utilize magnets, is based on the Compton 
scattering process \cite{Federici_EtAl_LettNuovoCim_v27_p339_1980}. 
In this process, a low-energy (eV) laser photon backscatters from an ultra-relativistic electron thus 
acquiring increase in the energy proportional to the squared Lorentz factor $\gamma=\E/mc^2$.
This method has been used for producing gamma-rays in a broad, $10^1$ keV -- $10^1$ MeV, energy range 
\cite{Rehman_EtAl_ANE_v105_p150_2017,Kraemer_EtAl-ScieRep_v8_p139_2018}.

The Compton scattering also occurs if the scatterer is an atomic (ionic) 
electron which moves being bound to a nucleus.
This phenomenon is behind the Gamma Factory proposal for CERN \cite{GammaFactory_CERN-Courier,Krasny:2018xxv}
that implies using a beam of ultra-relativistic ions in the backscattering process.
In this case, an ionic electron is resonantly excited by absorbing a laser photon.
The subsequent radiative de-excitation produces a gamma-photon. 

%%%%%%%%%%%%%%%%%%%%%%%%%%%%%%%%%%%%%%%%%%%%%%%%
\subsection{Brilliance and intensity of CUR
 \label{CUR-Bril}} 
 
The radiometric unit frequently used to compare different LS is \textit{brilliance}, $B$.  
It is defined in terms of the number of photons $\Delta N_{\omega}$ of frequency 
$\omega$ within the interval $[\omega-\Delta\omega/2,\omega+\Delta\omega/2]$ emitted in the cone 
$\Delta\Omega$ per unit time interval, unit source area, unit solid angle and per a 
bandwidth (BW) $\Delta\omega/\om$ \cite{SchmueserBook,RullhusenArtruDhez,EuroPhys}. 
To calculate this quantity is it necessary to know the beam electric current $I$,
transverse sizes $\sigma_{x,y}$ and angular divergences $\phi_{x,y}$ as well as
the divergence angle $\phi$ of the radiation and the 'size' $\sigma$ 
of the photon beam. 
Explicit expression for $B$ measured in
$\Bigl[\hbox{photons/s/mrad}^{2}\hbox{/mm}^{2}/0.1\,\%\,\hbox{BW}\Bigr]$ reads \cite{Kim2009}
%%%%%%%%%%%%%%
\begin{eqnarray}
B
=
{\Delta N_{\omega} \over 
10^{3}\, (\Delta\omega/\omega)\,(2\pi)^2\,\varepsilon_x\varepsilon_y}\,
{I \over e}\,,
\label{CUR_Brilliance:eq.01} %\label{B&B.7}
\end{eqnarray}
where $e$ is the elementary charge.
The quantities  
$\epsilon_{x,y}=\sqrt{\sigma^2+\sigma_{x,y}^2}\sqrt{\phi^2+\phi_{x,y}^2}$
are the total emittance of the photon source in the transverse directions
with $\phi=\sqrt{\Delta\Omega/2\pi}$ 
and $\sigma=\lambda/4\pi\phi$ being the `apparent' source size calculated in the diffraction limit
\cite{Kim1986NIM2}.
In~(\ref{CUR_Brilliance:eq.01}) $\sigma,\,\sigma_{x,y}$ 
are measured in millimeters, $\phi,\, \phi_{x,y}$ -- in milliradians.

The product $\Delta N_{\omega} I/e$ on the right-hand side of Eq. (\ref{CUR_Brilliance:eq.01}),
represents the number of photons per second (intensity) emitted in the cone $\Delta\Omega$ and 
frequency interval $\Delta\omega$ (see Eq. (\ref{Optimal_Length_CU:eq.01}) in Section \ref{Optimal_Length_CU}).
Using the peak value of the current, $I_{\max}$, one calculates
the \textit{peak brilliance}, $B_{\rm peak}$.

\begin{figure} [h]
\centering
\includegraphics[width=10cm,clip]{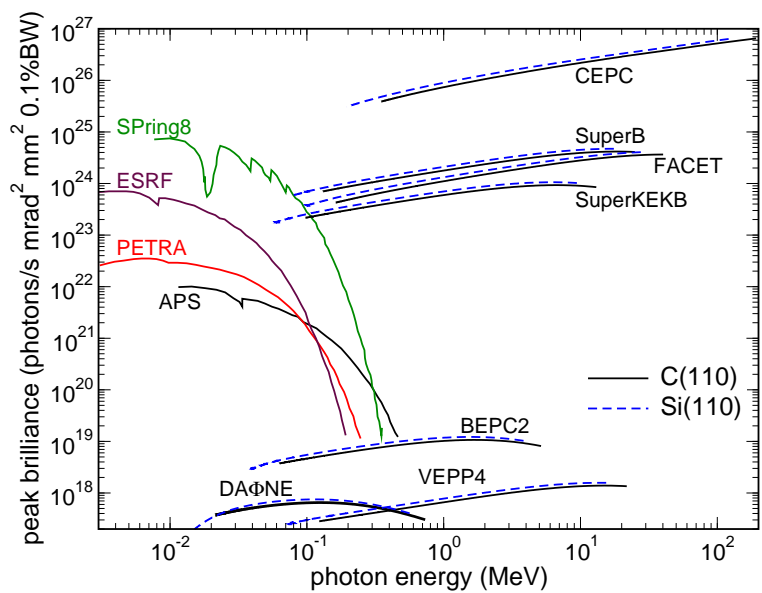}%{Bril_SR_vs_CU-2019_v01-mod.eps}
\caption{
Comparison of the peak brilliance available at several synchrotron radiation facilities
%SR facilities 
(APS, ESRF, PETRA, SPring8) with that for CUR from diamond(110)- and
Si(110)-based CUs for several positron beams listed in 
%Supplementary Table 1.
Table \ref{ep-beams-2018.Table} in \ref{Beams2018}.
The CUR data refer to the emission in the fundamental harmonic.
The data on APS (USA), ESRF (France), PETRA (DESY, Germany), SPring8 (Japan) are from 
\cite{Ayvazyan_EtAl_EPJD_v20_p149_2002,SchmueserBook}.
}
\label{Figure05.fig}%{Bril_SR_vs_CU-2019.fig}
\end{figure}

%%%%%%%%%%%%%%%%%%%%%%%%%%%%%%%%%%%%%%%%%%%%%%%%%
Let us compare the brilliance of CUR with that available at modern synchrotron facilities.
% SR facilities.
Figure \ref{Figure05.fig} presents the peak brilliance calculated for positron-based 
diamond(110) and Si(110) CUs and that for several synchrotrons.
%SR facilities. 
The CUR curves refer to \textit{the optimal parameters of CU} (see \ref{Optimal_Length_CU}), 
i.e. those which ensure the highest values of $B_{\rm peak}(\om)$ of CUR for each 
positron beam indicated.

%%%%%%%%%%%%%%%%%% 
%%%%%%%%%%%%%%%%%%%%%%%%%%%%%%%%%%%%%%

To be noted is that for the well-collimated intensive beams with small transverse sizes 
(SuperB, FACET, SuperKEK, CEPC) the peak brilliance of CUR in the photon energy range from $10^2$ keV  
to $10^2$ MeV (the corresponding wavelengths vary from $10^{-1}$ down to  $10^{-4}$ \AA{}) 
is comparable to (the case of SuperB, FACET and SuperKEK beams) or even higher 
(CEPC beam) than that achievable in conventional LS 
for much lower photon energies.

We stress that the values of bending amplitude and periods, which maximize the CUR brilliance
over broad range of photon energies, are accessible by means of modern technologies
(compare Figs. \ref{Figure08.fig} \& \ref{Figure09.fig} 
with Figs. \ref{Figure03.fig} \& \ref{Figure04.eps} in \ref{Beams2018}). 

%%%%%%%%%%%%%%%%% 
\begin{figure} [h]
\centering
\includegraphics[width=10cm,clip]{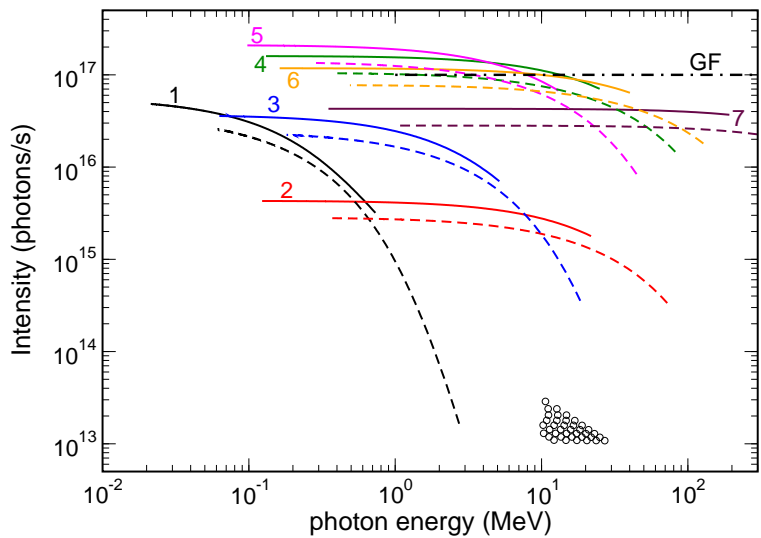}%{Nphot_sec_C_om1-3_2019_v04-mod.eps}
\caption{
Peak intensity (number of photons per second, $\Delta N_{\omega}\,I_{\max}/e$) of diamond(110)-based 
CUs calculated for positron beams at different facilities:
1 - DA$\Phi$NE, 2 - VEPP4M, 3 - BEPC-II, 4 - SuperB, 5 - SuperKEK, 6 - FACET-II, 7 - CEPC.
Solid and dashed lines correspond to the emission in the first and third harmonics,
respectively.
Open circles indicate the data on the laser-Compton backscattering \cite{Rehman_EtAl_ANE_v105_p150_2017}.
The horizontal dash-dotted line marks the intensity $10^{17}$ photon/s indicated in the 
Gamma Factory (GF) proposal for CERN \cite{Krasny:2018xxv}.
}
\label{Figure06.fig}%{Si-and-C110_intensity.fig}
\end{figure} 

The Gamma Factory proposal for CERN discusses a concept of the LS based on the resonant 
absorption of laser photons by the ultra-relativistic ions  \cite{GammaFactory_CERN-Courier,Krasny:2018xxv}.
It is expected that the intensity of the LS will be orders of magnitude
higher that the presently operating LS aiming at the values of $10^{17}$ photons/s 
in the gamma-ray domain $1 \leq E_{\rm ph} \leq 400$ MeV. 
To this end, it is instructive to compare the intensity 
of CUR with the quoted value as well as with the intensities currently achievable by means of the 
LS based on laser-Compton scattering from electron beam \cite{Rehman_EtAl_ANE_v105_p150_2017}.

Figure \ref{Figure06.fig} presents the peak intensities, $\Delta N_{\omega}\,I_{\max}/e$,
of the first (solid lines) and third (dashed lines) harmonics of CUR from diamond(110)-based 
CU with the optimized parameters (see Fig. \ref{Figure09.fig} in \ref{Optimal_Length_CU}).
Different curves correspond to different positron beams as specified in the caption. 
Most of the curves presented show orders of magnitude higher intensities in the photon energy range 
one to tens of MeV than that from the laser-Compton scattering LS (open circles).
Within the same photon energy interval the CUR intensity can be comparable with or even higher
(see the curves for the SuperB, SuperKEK and FACET-II beams)   
than the value predicted in the Gamma Factory proposal (marked with the horizontal dash-dotted line). 

Figures \ref{Figure05.fig} and \ref{Figure06.fig} demonstrate also the tunability of a CU-LS. 
For any positron beam with specified parameters the photon yield can be maximized (more generally, varied) 
over broad range of photon energies
by properly choosing parameters of the CU (bending amplitude and period, crystal, plane).   

%%%%%%%%%%%%%%%%%%%%%%%%%%%%%%%%%%%%%%%%%%%%%%%%
\subsection{Brilliance of superradiant CUR
 \label{CUL-Bril}} 

The radiation emitted in an undulator is coherent (at the harmonics frequencies) 
with respect to the number of periods, $N_{\rm u}$, but not with respect to the
emitters since the positions of the beam particles are not correlated.
As a result, the intensity of of radiation emitted in a certain direction is proportional to 
$N_{\rm u}^2$ and to the number of particles, $I_{\rm inc} \propto N_{\rm p} N_{\rm u}^2$
(the subscript `inc' stands for `incoherent').
In conventional undulators, $N_{\rm u}$ is on the level of $10^3\ldots 10^4$ \cite{SchmueserBook}, 
therefore, the enhancement due to the factor $N_{\rm u}^2$ is large making undulators a powerful source 
of spontaneous radiation.
However, the incoherence with respect to the number of the radiating particles causes a moderate (linear)  
increase in the radiated energy with the beam density.

%%%%%%%%%%%%%%%%%%%%%%%%%%%%%%%%%%%%%%%%%%%%%%%%%%%%%%%%%%%%%
More powerful and coherent radiation will be emitted by a beam in which position of the particles 
is modulated in the longitudinal direction with the period equal to integer multiple 
to the radiation wavelength $\lambda$.
In this case, the electromagnetic waves emitted by different particles have approximately the same phase.
Therefore, the amplitude of the emitted radiation is a coherent sum of the individual waves, so that the 
intensity becomes proportional to the number of particles squared,
$I_{\rm coh} \propto N_{\rm p}^2\, N_{\rm u}^2$ \cite{Bessonov-KvantElectron_v16_p1056_1986}.
Thus, the increase in the photon yield due to the beam pre-bunching (other terms used are `bunching' 
\cite{McNeilThompson_NaturePhotonics_v4_p814_2010} or `microbunching' \cite{Bostedt-EtAl_RMP_v88_015007_2016}) 
can reach orders of magnitudes relative to radiation by a non-modulated beam of the same density 
(see the data on $N_{\rm p}$  in Table \ref{ep-beams-2018.Table} in \ref{Beams2018}).
Following Ref. \cite{Gover-EtAl_RPM_v91_035003_2019} we use the term 'superradiant' to designate the 
coherent emission by a pre-bunched beam of particles.   

In what follows we assume that the beam is fully modulated at the crystal entrance.
The description on the methods of preparation of a pre-bunched beam with the parameters needed 
to amplify CUR one finds in \cite{Patent} and in Section 8.5 in Ref. \cite{ChannelingBook2014}.

For a pre-bunched beam, the intensity is sensitive not only to the shape of the trajectory but also to the 
relative positions of the particles along the undulator axis. 
In the course of beam propagation through the crystal these positions become random 
due to both the collisions with crystal atoms and the non-similarity of the 
channeling trajectories for different particles \cite{KKSG_stable_prop_2010}.
This leads to the beam demodulation and, as a result, to the loss of the superradiance effect.

For an unmodulated beam, the CU length $L$ is limited mainly by the dechanneling process. 
For a pre-bunched the demodulation becomes the phenomenon which imposes most restrictions 
on the parameters of a CU. 
In Ref. \cite{KKSG_stable_prop_2010} \textit{the demodulation length}, $L_{\rm dm}$, was introduced
to quantify the spatial scale at which a modulated beam becomes demodulated.
To preserve the modulation and to maintain the coherence of radiation the crystal length must be less 
than $L_{\rm dm}$ (see \ref{Demodulation} where essential details are summarized).

%%%%%%%%%%%%%%%%%% 
\begin{figure} [h]
\centering
\includegraphics[width=10cm,clip]{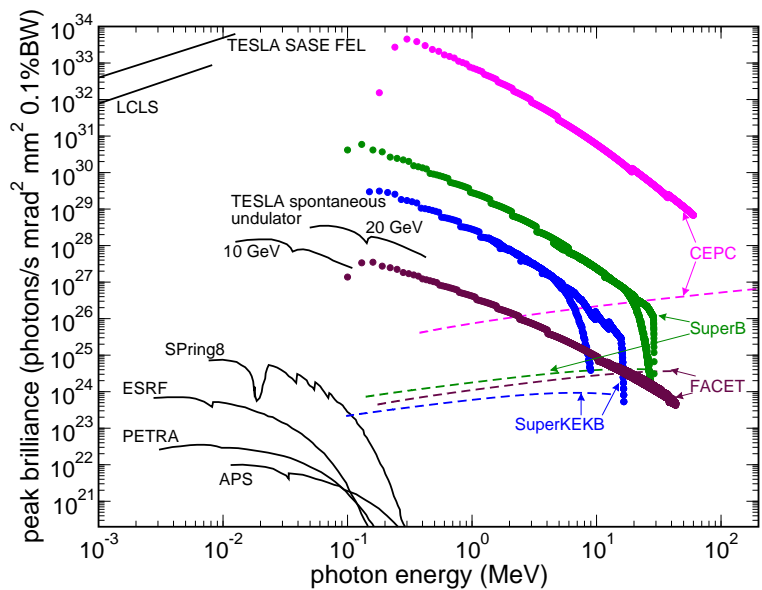}%{Bril_SR-FEL_vs_CU-CUL-2019_v08a-mod.eps}
\caption{
Peak brilliance of superradiant CUR (thick solid curves) and spontaneous CUR (dashed lines)  
from diamond(110) CUs calculated for the SuperKEKB, SuperB, FACET-II and CEPC positron beams 
versus modern synchrotrons, undulators and XFELs.
The data on the latter are taken from Ref. \cite{Ayvazyan_EtAl_EPJD_v20_p149_2002}.
}
\label{Figure07.fig}%{Bril_SR-FEL_vs_CU-CUL-2018.fig}
\end{figure}
% %%%%%%%%%%%%%%%%%%%%%%%%%%%%%%%%%%%%%

Quantitative analysis and numerical data on the parameters of a CU which maximize the brilliance of 
CUR in presence of the demodulation process is presented in Section \ref{SuperradiantCU}.
These data have been used to calculate the peak brilliance of the superradiant CUR. 

Figure \ref{Figure07.fig} illustrates a boost in peak brilliance due to the beam modulation.
Thick curves correspond to superradiant CUR calculated for fully modulated positron beams (as indicated) 
propagating in the channeling mode through diamond(110)-based CU.
In the photon energy range $10^{-1}\dots 10^1$ MeV the brilliance of superradiant CUR by orders of magnitudes
(up to 8 orders in the case of CEPC) exceeds that of the spontaneous CUR (dash-dotted curves) emitted 
by the random beams.
Remarkable feature is that the superradiant CUR brilliance can not only be much higher that the spontaneous 
emission from the state-of-the-art magnetic undulator (see the curves for the TESLA undulator) 
but also be comparable with the values achievable at the XFEL facilities (LCLC (Stanford) and TESLA SASE FEL)
which operate in much lower photon energy range.

%%%%%%%%%%%%%%%%%%%%%%%%%%%%%%%%%%%%%%%%%%%%%%%
\section{Discussion \label{Discussion}} 

Construction of novel CLSs is an extremely challenging task which constitutes a highly 
interdisciplinary field. 
To accomplish this task, a broad collaboration is needed of research groups with different but mutually complementary 
expertise, such as material science, nanotechnology, particle beam and accelerator physics, radiation physics, 
X-ray diffraction imaging, acoustics, solid state physics, structure determination, advanced computational modeling 
methods and algorithms, high-performance computing as well as industries specializing in manufacturing of 
crystalline structures and in design and construction of complete accelerator systems.

As a first step towards achieving the major breakthrough in the field, one can focus on practical 
realization of the CLS idea, i.e. elaboration of the key theoretical, experimental and technological 
aspects, demonstration of the device functionality and setting up of standards for the novel 
technology for construction of the CLSs aiming at their mass production in the future. 
Realization of this program implies a broad range of correlated and entangles activities including 
(i) Fabrication of linear, bent and periodically bent crystalline structures with lattice 
quality necessary for delivering pre-defined bending parameters within the ranges indicated in 
Figure \ref{Figure03.fig} in \ref{Beams2018};
(ii) Advanced control of the lattice quality by means of the highest quality non-destructive 
X-ray diffraction techniques.
The same techniques to be applied to detect possible structural modification following particle irradiation;
(iii) Validation of functionality of the manufactured structures through experiments with high-quality  
(low energy spread, low emittance, high particle density and current) beams of ultra-relativistic electrons and 
positrons with $\E=10^{-1}- 10^1$ GeV, including an authoritative study of the structure 
sustainability with respect to beam intensity, as well as explicit experimental characterization 
of the emission spectra;
(iv) Advance in computational and numerical methods for multiscale modeling of nanostructured materials 
with extremely high, reliable levels of prediction (from atomistic to mesoscopic scale), 
of particle propagation, of irradiation-induced solid state effects, and for calculation of spectral-angular 
distribution of emitted radiation and for modeling \cite{MBN_Explorer_Paper}. 
Ultimately, this will enable better experimental planning and minimisation of experimental costs. 
The knowledge gained the studies (i)-(v) will provide CLSs prototypes and a roadmap for practical 
implementation by CLS system manufacturers and accelerator laboratories/users worldwide.

Sub-angstrom wavelength powerful and tunable CLSs will have a broad range of exciting 
potential applications.

A micron-sized narrow photon beam may be used in {\it cancer therapy}.
This would greatly improve the precision and effectiveness of the therapy for the destruction of
tumour by collimated radiation.
Furthermore, it would allow delicate operations to be performed in close vicinity of vital organs.
Taking into account the experience gained to date in the field of radio-therapy, one can expect that practical 
manipulations with micro-sized beams will become active soon after the novel LSs become available.

Gamma-rays induce {\it nuclear reactions by photo-transmutation}.
For instance, in the experiment of Ref. \cite{Ledingham-EtAl_Science_v300_p1107_2003} a long-lived isotope 
was converted into a short-lived one by irradiation with a gamma-ray bremsstrahlung pulse.
However, the intensity of bremsstrahlung is orders of magnitudes less than of CUR.
Moreover, to increase the effectiveness of the photo-transmutation process is it
desirable to use photons whose energy is in resonance with the transition energies
in the irradiated nucleus \cite{Rehman_EtAl_ANE_v105_p150_2017}.
By varying the CU parameters one can tune the energy of CUR to values needed
to induce the transmutation process in various isotopes.
This opens the possibility for a {\it novel technology for disposing of nuclear waste}.
Photo-transmutation can also be used to produce {\it medical isotopes}.
Another possible application of the CU-LSs concerns {\it photo-induced nuclear
fission} when a heavy nucleus is split into two or more fragments due to the irradiation
with gamma-quanta whose energy is tuned to match the transition energy between the nuclear states.
% ,
This process can be used in a new type of nuclear reactor -- the photo-nuclear reactor.
The production of Positron Emission Tomography (PET) isotopes will be very favourable application, exploiting 
the $(\gamma,n)$ reaction in the region of the giant dipole resonance (typically 20-40 MeV).
The PET isotopes can be used directly for medial PET and for Positron Emission Particle Tracking experiments.
Powerful monochromatic radiation within the MeV range can be used as an alternative source for producing beams 
of MeV protons by focusing a photon pulse on to a solid target 
\cite{Ledingham-EtAl_Science_v300_p1107_2003,Ledingham-EtAl_EurPhysNews_v33_p120_2002}.
Such protons can induce nuclear reactions in materials producing, in particular, 
light isotopes which serve as positron emitters to be used in PET.
Irradiation by hard X-ray strongly decreases the effects of natural surface tension of water 
\cite{Weon-EtAl_PRL_v100_217403_2008}.
The possibility to tune the surface tension by the irradiation can be exploited to study
the many phenomena affected by this parameter in physics, chemistry, and biology such as, 
for example, the tendency of oil and water to segregate.

%%%%%%%%%%%%%%%%%%%%%%%%%%%%%%%%%%%%%%%%%%%%%%%
\section{Conclusion \label{Conclusion}} 

The exemplary case study of a tunable CU-based LS considered in the paper has demonstrated that 
peak brilliance of CUR emitted in the photon energy range $10^2$ keV up to $10^2$ MeV by currently 
available (or planned to be available in near future) positron beams channeling in 
%PBCs 
periodically bent crystals is comparable 
to or even higher than that achievable in conventional synchrotrons in the much lower photon energy range. 
Intensity of CUR greatly exceeds the values provided by LSs based on Compton scattering 
and can be made higher than the values predicted in the Gamma Factory proposal in CERN. 
By propagating a pre-bunched beam the brilliance in the energy range $10^2$ keV up to $10^1$ MeV can be 
boosted by orders of magnitude reaching the values of spontaneous emission from the state-of-the-art magnetic 
undulators and being comparable with the values achievable at the XFEL facilities which operate  
in much lower photon energy range.  
Important is that by tuning the bending amplitude and period one can maximize brilliance for given parameters of 
a positron beam and/or chosen type of a crystalline medium.
Last but not least, it is worth to mention that the size and the cost of CLSs are orders of magnitude 
less than that of modern LSs based on the permanent magnets. 
This opens many practical possibilities for the efficient generation of gamma-rays with various intensities 
and in various ranges of wavelength by means of the CLSs on the existing and newly constructed beam-lines.

Though we expect that, as a rule, the highest values of brilliance can be reached in CU-based LSs (or, in those 
based on stacks of CUs) the analysis similar to the one presented can be carried out for other types 
of CLSs based on linear and bent crystals. 
This will allow one to make an optimal choice of the crystalline target and the CLS type 
to be used in a particular experimental environment or/and to tune the parameters of the emitted 
radiation matching them to the needs of a particular application.

The case study presented has been focused on the positron beams, which have a clear advantage since the dechanneling 
length of positrons is order of magnitude larger than that of electrons of the same energy.
This allows one to use thicker crystals in channeling experiments with positrons thus enhancing the 
photon yield.
Nevertheless, experimental studies of CLSs with electron beams are worth to be carried out.
Indeed, high-quality electron beams of energies starting from sub-GeV range and onward are more 
available than their positron counterparts. 
Therefore, these laboratories provide more options for the design, assembly and practical 
implementation of a full suite of correlated experimental facilities needed for operational 
realization and exploitation of the novel CLSs. 
In this connection we note that in the course of channeling experiments at the Mainz Microtron facility 
with $\E=190-855$ MeV electrons propagating in various CUs, which have been carried out over the last 
decade within the frameworks of several EU-supported collaborative projects (FP6-PECU, FP7-CUTE, H2020-PEARL),   
a unique experience has been gained.
This experience has ascertained that the fundamental importance of the quality of 
% PBCs, 
periodically bent crystals, which, in turn is 
based on the cutting-edge technologies used to manufacture the crystalline structures, of modern techniques 
for non-destructive characterization of the samples, of the necessity of using advanced computational 
methods for numerical modeling of a variety of phenomena involved.
On the basis of this experience the bottlenecks on the way to practical realization of 
the CLSs concept have been established. 

To quantify the scale of the impact within Europe and worldwide which the development of radically 
novel CLSs might have, we can draw historical parallels with synchrotrons, optical lasers and FELs. 
In each of these technologies there was a significant time lag between the formulation of a pioneering idea, 
its practical realization and follow-up industrial exploitation. 
However, each of these inventions has subsequently launched multi-billion dollar industries. 
The implementation of CLS, operating in the photon energy range up to hundreds of MeV, is expected 
to lead to a similar advance and CLSs have the potential to become the new synchrotrons and lasers of the mid 
to late 21st century, stimulating many applications in basic sciences, technology and medicine. 
The development of CLS will therefore herald a new age in physics, chemistry and biology.

%%%%%%%%%%%%%%%%%%%%%%%%%%%%%%%%%%%%%%%%%%%%%%%%%
\ack

We acknowledge support by the European Commission through the N-LIGHT Project within the H2020-MSCA-RISE-2019 call (GA 872196).
The work was also supported by Deutsche Forschungsgemeinschaft (Project No. 413220201).

% \vspace*{1cm}
% \noindent
% \textbf{Author contributions}\\
% A.V.K. \dots
% \\
% A.V.S. \dots

\appendix

%%%%%%%%%%%%%%%%%%%%%%%%%%%%%%%%%%%%%%%%%%%%%%%%%%%%%%%%%%%%%%%%%
\section{Beam Parameters
\label{Beams2018}}

The data on positron and electron beams energy $\E$, bunch length $\Lb$, number of particles per 
bunch $\calN$, beam sizes $\sigma_{x,y}$ and divergences $\phi_{x,y}$ (the subscripts $x,y$ refer 
to the horizontal and vertical dimensions, respectively), volume density $n$,
and peak current $I_{\max}$ are summarized in Table \ref{ep-beams-2018.Table}.
The table compiles the data for the following facilities:
VEPP4M (Russia), BEPCII (China), DA$\Phi$NE (Italy),  SuperKEKB (Japan) 
\cite{ParticleDataGroup2018},
SLAC (the FACET-II beams, Ref. \cite{FACETII_Conceptual_Design_Rep-2015}),
SuperB (Italy) \cite{ParticleDataGroup2010}, and 
CEPC (China) \cite{CEPC-Design-Report_arXiv1809-00285_2018}.
Note that the SuperB data are absent in the latest review by Particle Data Group 
\cite{ParticleDataGroup2018} since its construction was canceled \cite{SuperB-Cancelled}.

%%%%%%%%%%%%%%%%%%%%%%%%%%%
\begin{table}[h]
\hspace*{-2cm}\caption{
Parameters of positron ('p') and electron ('e') beams:
beam energy, $\E$,
bunch length, $\Lb$,
number of particles per bunch, $\calN$,
beam size, $\sigma_{x,y}$,
beam divergence $\phi_{x,y}$,
volume density $n=\calN/(\pi \sigma_{x}\sigma_{y}\Lb)$ of particles in the bunch,
peak current $I_{\max}=e\calN c/\Lb$.
In the cells with no explicit reference to either 'e' or 'p' the data refer to both modalities.
}
\footnotesize\rm
\begin{tabular}{@{}lllllllll}
\br
 Facility        &VEPP4M&BEPCII &DA$\Phi$NE& SuperKEKB       & SuperB              & FACET-II& CEPC  \\
 Ref.            &\cite{ParticleDataGroup2018}    
                        &\cite{ParticleDataGroup2018}
                                  & \cite{ParticleDataGroup2018}
                                             & \cite{ParticleDataGroup2018} 
                                                         & \cite{ParticleDataGroup2010,SuperB-Cancelled}
                                                         &\cite{FACETII_Conceptual_Design_Rep-2015}
                                                         &\cite{CEPC-Design-Report_arXiv1809-00285_2018}\\
\br
$\E$ (GeV)         &6    &1.9-2.3& 0.51   &{\rm p}: 4     &{\rm p}: 6.7  & 10    & 45.5 \\
                   &     &       &        &{\rm e}: 7     &{\rm e}: 4.2  &       &      \\
\mr 
$\calN$            &15   & 3.8   &{\rm p}: 2.1    &{\rm p}: 9.04  &{\rm p}: 6.5  &{\rm p}: 0.375 &  8   \\ 
(units 10$^{10}$)  &     &       &{\rm e}: 3.2    &{\rm e}: 6.53  &{\rm e}: 5.1  &{\rm e}: 0.438 &       \\ 
\mr  
$\Lb$ (cm)         &5    & 1.2   & 1-2    &{\rm p}: 0.6   & 0.5  &{\rm p}: 0.00076&  0.85\\
                   &     &       &        &{\rm e}: 0.5   &      &{\rm e}: 0.00011&   \\
\mr
$\sigma_x$ ($\mu$m)&1000 & 347   & 260    &{\rm p}: 10    &  8   &{\rm p}: 10.1  & 6    \\
                   &     &       &        &{\rm e}: 11    &  8   &{\rm e}: 5.5   &      \\
$\sigma_y$ ($\mu$m)&30   & 4.5   & 4.8    &{\rm p}: 0.048 & 0.04 &{\rm p}: 7.3   & 0.04 \\
                   &     &       &        &{\rm e}: 0.062 &      &{\rm e}: 5.9   &      \\
\mr
$\phi_x$ (mrad)    &0.2  & 0.35  &   1    &{\rm p}: 0.32  &{\rm p}: 0.250&{\rm p}: 0.178 & 0.03 \\
                   &     &       &        &{\rm e}: 0.42  &{\rm e}: 0.313&{\rm e}: 0.073 &    \\
$\phi_y$ (mrad)    &0.67 & 0.35  & 0.54   &{\rm p}: 0.18  &{\rm p}: 0.125&{\rm p}: 0.044 & 0.04 \\
                   &     &       &        &{\rm e}: 0.21  &{\rm e}: 0.150&{\rm e}: 0.044 &      \\
\mr
$I_{max}$ (A)      &144  & 152   &{\rm p}: 50-100 &{\rm p}: 723   &{\rm p}: 624  &{\rm p}: 12.1$\times10^{3}$  & 452  \\
                   &     &       &{\rm e}: 77-154 &{\rm e}: 627   &{\rm e}: 490  &{\rm e}: 75.5$\times10^{3}$  &      \\
\mr
 $n$              
 ($10^{13}$cm$^{-3}$)
                  &3.2   & 65    &{\rm p}:  54     &{\rm p}:$1.0\times10^{6}$&{\rm p}:$1.3\times10^{6}$
                                                                                &{\rm p}: $2\times10^{5}$
                                                                                  &$1.25\times10^{6}$\\
                  &      &       &{\rm e}:  82     &{\rm e}:$0.6\times10^{6}$&{\rm e}:$1.0\times10^{6}$
                                                                                 &{\rm e}: $3.9\times10^{6}$
                                                                                   &                  \\
\br
\end{tabular}
\label{ep-beams-2018.Table}
\end{table}

%%%%%%%%%%%%%%%%%%%%%%%%%%%%%%%%%%%%%%%%%%%%%%%
\section{Optimal Length of a CU \label{Optimal_Length_CU}}

With account for the dechanneling and the photon attenuation, the number of photons $\Delta N_{\omega_n}$ of the frequency 
within the interval $\Bigl[\omega_n-\Delta\omega_n/2,\omega_n+\Delta\omega_n/2\Bigr]$ emitted in the forward direction 
within the cone $\Delta\Omega_n$ by a projectile in a CU is given by the following expression
(see Refs. \cite{SPIE1,ChannelingBook2014} for the details):
\begin{eqnarray}
\Delta N_{\omega_n}
=
\textcolor{black}{\calA(C)}\,
{4\pi\alpha\, n \zeta}
\left[
J_{{n-1 \over 2}}(n\zeta) - J_{{n+1 \over 2}}(n\zeta)
\right]^2
\Neff(\Nd; x,{\kappa_{\mathrm{d}}})\,
{\Delta \omega_n \over \omega_n} ,
\label{Optimal_Length_CU:eq.01} %{CUR_Brilliance:eq.02} 
\end{eqnarray}
where $\zeta = K^2/(4+2K^2)$, $J_{\nu}(n\zeta)$ is the Bessel function and $K=2\pi\gamma a/\lamu$ is the
undulator parameter.
The subscript $n$ enumerates the harmonics of CUR. 
The frequency $\om_n=n\om_1$ of the $n$-th harmonic is expressed in terms of the fundamental harmonic given by
\begin{eqnarray}
\om_1
= 
{ 2 \gamma^2 \over 1 + K^2/2}{2\pi c\over \lamu}\,.
\label{Optimal_Length_CU:eq.02} %{alpha-beta:eq.08}
\end{eqnarray}
 \textcolor{black}{
 The quantity $\calA$ stands for the channel acceptance, which is defined as a fraction of the incident particles
 captured into the channeling mode at the crystal entrance 
 (another term used is surface transmission, see e.g. Ref. \cite{Wistisen_etal:PR-AB_19_071001_2016}).
 }
 
\textcolor{black}{Apart from the factor $\calA$,}
%The 
the difference between~(\ref{Optimal_Length_CU:eq.01}) and the formula for an ideal undulator 
(see, e.g., \cite{Kim2009}) is that the number of undulator periods $\Nu$, which enters the latter, 
is substituted with the \textit{effective number of periods}, $\Neff(\Nd,x,\kappa_{\d})\equiv \Neff$, 
which depends on the number of periods within the dechanneling length, $\Nd=\Ld/\lamu$, and on the ratios 
$x=\Ld/\La$  and $\kappa_{\d}=L/\Ld$ where $\Ld$ denotes the dechanneling length and
$\La$ is the photon attenuation length.
The effective number of periods is given by \cite{SPIE1,ChannelingBook2014}:
%%%%%%%%%%%%%%%%%%%%
\begin{eqnarray}
\fl
\Neff
=
\frac{4 \Nd}{x \kappa_{\d}}
\left[
{x \ee^{-x\kappa_{\d}} \over (1-x)(2-x)}
-
{\ee^{-\kappa_{\d}} \over 1-x}
+
{2\ee^{-(2+x)\kappa_{\d}/2} \over 2-x}
\right]
\sqrt{
1
+\kappa_{\d}^2 {(x-1)^2 +1 \over 4\pi^2}
}\,.
\label{Optimal_Length_CU:eq.03} %{CUR_Brilliance:eq.03} 
\end{eqnarray}
In the limit $\Ld, \La \to \infty$, i.e. when  the dechanneling and the attenuation are neglected, 
$\Neff \to \Nu = L/\lamu$, as it must be in the case of an ideal undulator.
In this case one can, in principle, increase infinitely the number of periods by
considering larger values of the undulator length $L$.
This will lead to the increase of the number of photons and the brilliance since these
quantities are proportional to $\Nu$.
The limitations on the values of $L$ and $\Nu$  are mainly of a technological nature.

The situation is different for a CU, where the number of channeling particles and the number of
photons, which can emerge from the crystal, decrease with the growth of $L$.
It is seen from (\ref{Optimal_Length_CU:eq.03}), that in the limit $L \to \infty$ the parameters 
$\kappa_{\d}$ and $x \kappa_{\d}=L/\La$ also become infinitely large leading to $\Neff \to 0$.
This result is quite clear, since in this limit $L \gg \La$ so that all emitted photons are 
absorbed inside the crystal.
Another formal (and physically trivial) fact is that 
$\Neff=0$ also for a zero-length undulator $L=0$.
Vanishing of a positively-defined function $\Neff(\Nd,x,\kappa_{\d})$ at two extreme boundaries suggests 
that there is a length $\overline{L}(x)$ which corresponds to the maximum value of the function.
 
To define the value of $\overline{L}(x)$ or, what is equivalent, of the quantity 
$\overline{\kappa}_\d(x)=\overline{L}(x)/\Ld$, one carries out the derivative of 
$f(x,\kappa_\d)$ with respect to $\kappa_\d$ and equalizes it to zero.
The analysis of the resulting equation shows that for each value of $x=\Ld/\La\geq 0$ there is only
one root $\overline{\kappa}_\d$.
Hence, the equation defines, in an inexplicit form, a single-valued function 
$\overline{\kappa}_\d(x)=\overline{L}(x)/\Ld$ which ensures the maximum of 
$N_{\rm eff}(x,\kappa_{\rm d})$ for given $\La$, $\Ld$ and $\lamu$.

Note that the crystal length enters Eq.\,(\ref{Optimal_Length_CU:eq.01}) 
only via the ratio ${\kappa_{\mathrm{d}}}$.
It was shown \cite{SPIE1,ChannelingBook2014} that the quantity $\overline{L}(x)$ ensures the highest 
values  of the number of photons $\Delta N_{\omega_n}$ and of the brilliance $B_n$ of the CUR.
Therefore, $\overline{L}(x)$ can be called the  \textit{optimal length} that corresponds to 
a given value of the ratio  $x=\Ld/\La$.
 
The following multi-step procedure has been adopted to calculate the highest brilliance  of CUR.
 
\begin{itemize}
%%%%%
\item
 \textit{Fix crystal and crystallographic direction.} 
 In the current paper we have focused on the (110) planar channels in diamond and silicon crystals, 
 which are commonly used in channeling experiments.
We note that other crystals/channels, available or/and studied experimentally, can also be considered
 \cite{KSG_review_1999,KKSG_demodulation,SytovEtAl_EPJC_v77_901_2017}.
 
%%%%%
\item
 \textit{Fix parameters of the positron beam:} energy $\E$, sizes $\sigma_{x,y}$ and 
 divergence $\phi_{x,y}$, peak beam current $I_{\max}$.

%%%%%
\item 
\textit{Scan over photon energy $\om$.}
For each $\om$ value:

% \begin{itemize}
\begin{enumerate}
%%%%%%
\item Determine the attenuation length $\La(\om)$ (for the photon energies above
  1 keV the data are compiled in Ref. \cite{Hubbel}).
%%%%%%
\item Scan over $a$ and $\lamu$  consistent with the stable channeling condition
  \cite{KSG1998,KSG_review_1999}:
\begin{eqnarray}
C = 4\pi^2 {a\over \lambda_{\rm u}^2} {\E \over \dUmax}
< 1\,.
\label{Optimal_Length_CU:eq.04} 
\end{eqnarray}
The bending parameter $C$ is defined as the ratio $F_{\rm cf}/\dUmax$
where $F_{\rm cf}\approx \E/R$ is the centrifugal force in a channel bent with 
curvature radius $R$ and $\dUmax$ is the maximum force due to the interplanar potential.
Channeling motion in the bent crystal is possible if $C<1$.
In a 
%PBC, 
periodically bent crystal, the bending radius in the points of maximum curvature equals to
$\lambda_{\rm u}^2/4\pi^2 a$ which explains the right-hand side of 
(\ref{Optimal_Length_CU:eq.04}).

%%%%%
\item Determine dechanneling length $L_d(C)$.

The data on the dechanneling length can be extracted (when available) 
from the experiments \cite{BackeEtAl_JINST_v13_C04022_2018,WienandsEtAl_PRL_2015} 
or obtained by means of highly accurate numerical simulation of the channeling process
\cite{MBN_ChannelingPaper_2013,ChannelingBook2014,MBNExplorer_Book}.
For positrons, a very good estimation for $L_d(C)$ can be obtained by means of the following 
formulae \cite{BiryukovChesnokovKotovBook,ChannelingBook2014}:
\begin{eqnarray}
\begin{array}{ll}
\Ld(C) = (1-C)^2\Ld(0),
\label{Optimal_Length_CU:eq.05} % {Ld:eq.01}
\\
\displaystyle
\Ld(0)
=
{256 \over 9\pi^2}\,
{\aTF d \over r_0 m_ec^2}\,
{\E \over \Lambda}
\label{Optimal_Length_CU:eq.06} %{Ld:eq.02}
\end{array}
\end{eqnarray}
Here $\Ld(0)$  is the dechanneling length in the straight channel, $r_0$ cm is the classical electron radius,
$Z$ and $\aTF$ are, respectively, the atomic number and the Thomas-Fermi radius of the constituent atom, 
$\Lambda =13.55 + 0.5\ln(\E\mbox{[GeV]}) - 0.9\ln(Z)$ is the Coulomb logarithm.

%%%%%
\item Determine the maximum value of $\Neff$ and the optimal length $\overline{L}$ .

%%%%%
\item 
\textcolor{black}
{Determine the channel acceptance.}

\textcolor{black}{
The acceptance $\calA(C)$ of a bent channel can be estimated as follows \cite{BiryukovChesnokovKotovBook}:
\begin{eqnarray}
\calA(C) = (1-C) \, \calA_0.
\label{Optimal_Length_CU:eq.07} 
\end{eqnarray}
Here $\calA_0 = 1 - 2u_T/d$ ($u_T$ is the amplitude of thermal vibrations of the crystal atoms) is the acceptance of the straight channel.
}

%%%%%
\item Substituting the quantities obtained into Eq. (\ref{Optimal_Length_CU:eq.01}) and Eq.(\ref{CUR_Brilliance:eq.01}) 
in the main text one calculates the highest available peak brilliance $B_{\rm peak}(\om)$.
\end{enumerate}
% \end{itemize}

\end{itemize}

\textcolor{black}{
As formulated, the items (iii)-(vi) listed above are applicable for a fully collimated positron beam with zero divergence.
In reality, the beams have non-zero divergence}
\textcolor{black}{$\phi$, see Table \ref{ep-beams-2018.Table}, so that only a fraction $\xi$
of the beam particles gets accepted into the critical angle $\Theta_{\rm L}$ for channeling.
To estimate this fraction we assume the normal distribution of the beam particles with respect to the incident angle and
calculate $\xi$ as follows:
\begin{eqnarray}
\xi = (2\pi\phi^2)^{-1/2}\int_{-\Theta_{\rm L}}^{\Theta_{\rm L}} \exp\left(- {\theta^2 \over 2\phi^2}\right) \d \theta.
\label{Optimal_Length_CU:eq.08} 
\end{eqnarray}
The values of $\xi$ calculated for the beams listed in Table \ref{ep-beams-2018.Table} are presented in Table \ref{ep-beams-2018.Table-2}.
For each beam, Lindhard's critical angle $\Theta_{\rm L}=\sqrt{2U_0/\E}$ is estimated using the value $U_0=20$ eV 
(which corresponds, approximately, to the interplanar potential depth in Si(110) and diamond(110))
and the indicated values of the beam divergence is calculated as $\phi = \min[\phi_x,\phi_y]$.
}

\textcolor{black}{
To account for the non-zero divergence one multiplies the value $B_{\rm peak}(\om)$, 
calculated as described above, by the factor $\xi$.
}

%%%%%%%%%%%%%%%%%%%%%%%%%%%
\begin{table}[h]
\hspace*{-2cm}\caption{
\textcolor{black}{
Fraction $\xi$ of the beams particles with incident angle less than Lindhard's critical angle $\Theta_{\rm L}$ 
(in mrad).
For each beam indicated the parameter $\phi$ (in mrad) stands for the minimum of two divergences $\phi_{x}$ and $\phi_{y}$, 
see Table \ref{ep-beams-2018.Table}.
}
}
\footnotesize\rm
\begin{tabular}{@{}lllllllll}
\br
 Facility             & VEPP4M & BEPCII & DA$\Phi$NE & SuperKEKB & SuperB & FACET-II & CEPC   \\
\br
$\phi$                & 0.2    & 0.35   &  0.54      & 0.18      & 0.125  & 0.044    & 0.03   \\
\mr  
$\Theta_{\rm L}$      & 0.08   & 0.14   & 0.28       & 0.1       & 0.08   & 0.063    & 0.03   \\
$\xi$                 & 0.31   & 0.31   & 0.40       & 0.42      & 0.48   & 0.85     & 0.67   \\
\br
\end{tabular}
\label{ep-beams-2018.Table-2}
\end{table}

\begin{figure} [h]
\centering
\includegraphics[scale=0.45,clip]{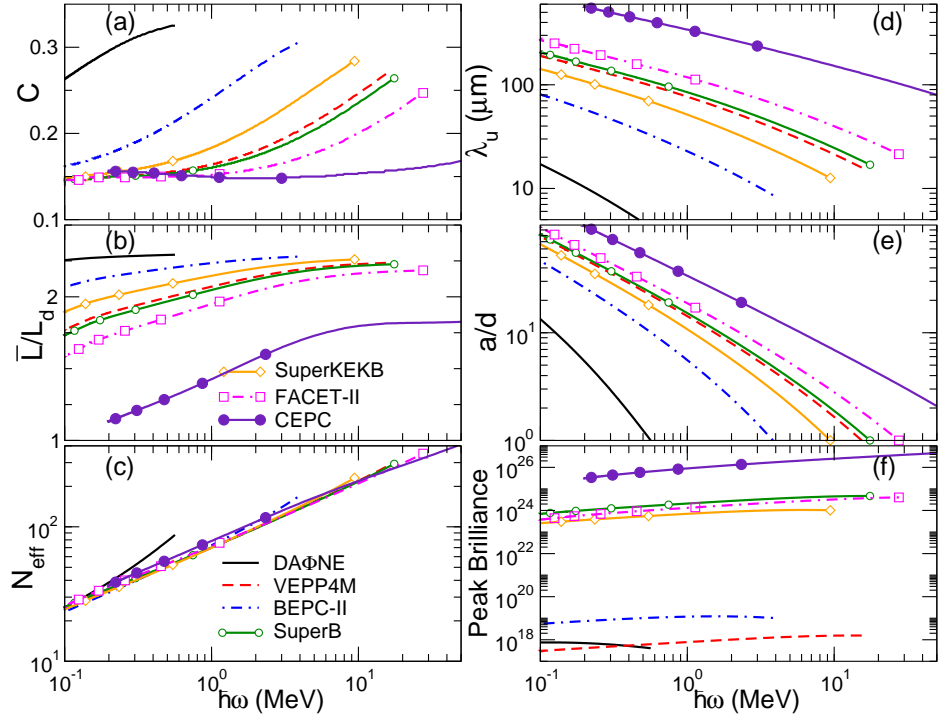}%{Si110_parameters_v06.eps}
\caption{
Parameters of the silicon(110)-based CU, - $C$, $\lamu$, $a$ (measured in the interplanar
distances $d$), $\Neff$, $\overline{L}$ (measured in the units of $\Ld(C)$), 
that ensure the highest peak brilliance $B_{\rm peak}(\om)$, graph (f).
Different curves correspond to several currently achievable
positron beams as indicated in the legend (see also Table  \ref{ep-beams-2018.Table}).}
\label{Figure08.fig}%{Si110_parameters.fig}
\end{figure}

\begin{figure} [h]
\centering
\includegraphics[scale=0.45,clip]{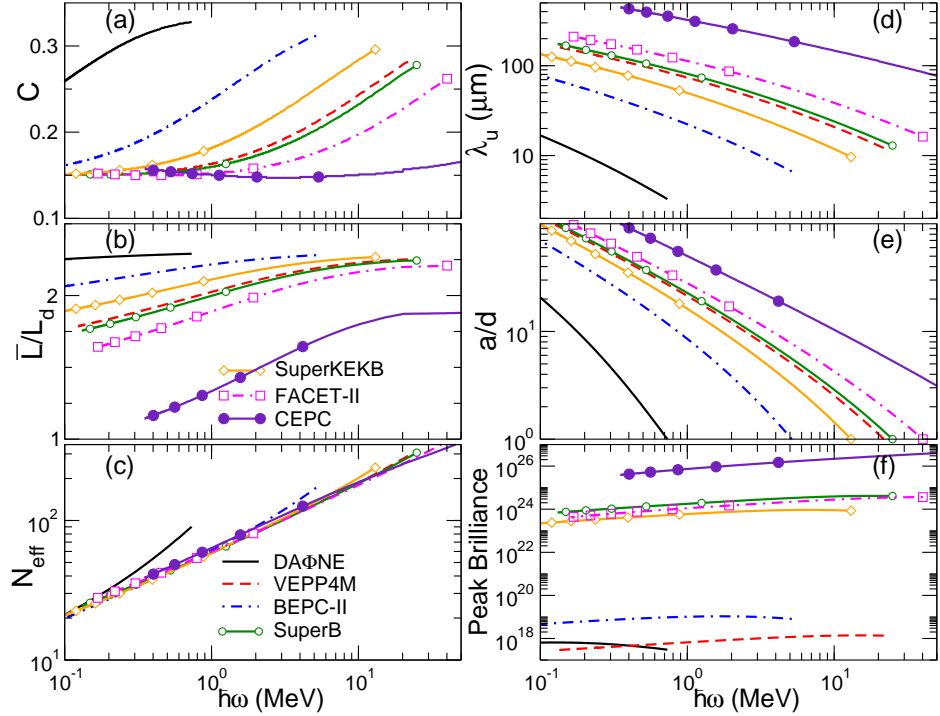}%{C110_parameters_v06.eps}
\caption{
Same as in Fig. \ref{Figure08.fig} but for the diamond(110)-based CU.}
\label{Figure09.fig}%{C110_parameters.fig}
\end{figure}

Figures \ref{Figure08.fig} and \ref{Figure09.fig} show the results of 
calculations performed for silicon(110)- and diamond(110)-based CU using and
for the positron beams specified in Table  \ref{ep-beams-2018.Table}.
The dependences presented were obtained by maximizing the brilliance
of CUR emitted in the fundamental harmonic.
It is seen, that within the range of moderate values of the bending amplitude ($a/d$ varies from several units 
up to several tens, graphs (e); $d=1.26$ and 1.92 \AA{} for the diamond and silicon crystals, respectively) 
it is possible to construct a CU with sufficiently large number of effective periods, 
$\Neff\approx 10\dots 100$, graphs (c).
These values correspond to the range of undulator periods $\lamu \approx 10^1 \dots 10^2$\,$\mu$m (graphs (d)) 
which is achievable by different methods of preparation of
periodically-bent crystalline structures, see Section \ref{Feasibility-Methods}.
It is seen from Figs.\,\ref{Figure08.fig} and \ref{Figure09.fig} that out of all calculated quantities the 
peak brilliance $B_{\rm peak}(\om)$, graphs (f), is the most sensitive to
the parameters of the positron beam.
The variation in the magnitude of $B_{\rm peak}(\om)$ is over six orders of magnitude, from $\sim 10^{18}$ 
up to $\sim10^{26}\mbox{photons/s/mrad}^{2}/\hbox{mm}^{2}/0.1\,\%\,\hbox{BW}$ (compare the DA$\Phi$NE and CEPC curves).

%%%%%%%%%%%%%%%%%% 
% %%%%%%%%%%%%%%%%%%%%%%%%%%%%%%%%%%%%%
%%%%%%%%%%%%%%%%%% 
% %%%%%%%%%%%%%%%%%%%%%%%%%%%%%%%%%%%%%

%%%%%%%%%%%%%%%%%%%%%%%%%%%%%%%%%%%%%%%%%%%%%%%%%%%%
\section{Beam Demodulation in CU} \label{Demodulation}

In a CU, a channeling particle, while moving along the channel centerline, undergoes two other types of motion 
in the transverse directions with respect to the CU axis $z$. 
First, there are channeling oscillations along the $y$ direction perpendicular to the crystallographic planes. 
Second, the particle moves along the planes (the $x$ direction).
To be noted is that different particles have different
(i) amplitudes $a_{ch}$ of the channeling oscillations, 
and 
(ii)  momenta $p_x$ in the $(xz)$ plane due to the distribution in the transverse energy of the beam particles as 
well as the result of multiple scattering from crystal atoms.
Therefore, even if the speed of all particles along their trajectories is the same,
the difference in $a_{\rm ch}$ or/and in $p_x$ leads to different values of the
velocities with which the particles move along the undulator axis.
As a result, the beam loses its modulation while propagating through the crystal.

For an unmodulated beam, the CU length $L$ is limited mainly 
by the dechanneling process. 
A dechanneled particle does not follow the periodic shape of the channel, 
and, thus, does not contribute to the CUR spectrum.
Hence, it is reasonable to estimate $L$ on the level of several dechanneling lengths $\Ld$ (see panels (b) 
in Figs. \ref{Figure08.fig} and \ref{Figure09.fig}).
Longer crystals would attenuate rather then produce the radiation.
Since the intensity of CUR is proportional to the undulator length squared, the dechanneling length and 
the attenuation length are the main restricting factors (see Section \ref{Optimal_Length_CU}) which must 
be accounted for.

For a modulated beam, the intensity is sensitive not only to the shape of the 
trajectory but also to the relative positions of the particles along the undulator axis.
If these positions become random because of the beam demodulation, the coherence of
CUR is lost even for the channeled particles.
Hence, the demodulation becomes the phenomenon which imposes most restrictions on the
parameters of a CU.

In Ref. \cite{KKSG_stable_prop_2010} an important quantity, -- \textit{the demodulation length}, was introduced.
It represents the characteristic scale of the penetration depth at which a
modulated beam of channeling particles becomes demodulated.
Within the framework of the approach developed in the cited papers the demodulation length $L_{\rm dm}$ is related 
to the dechanneling length $\Ld(C)$ in a bent channel:
\begin{equation}
\Ldm 
= 
{ \Ld(C)  \over \alpha(\xi) + \sqrt{\xi}/j_{0,1}}\,.
\label{Demodulation:eq.01}
\end{equation}
Here $j_{0,1}=2.4048\dots$ is the first root of the Bessel function $J_0(x)$. 
The dimensionless parameter $\xi$ is expressed in terms of the emitted radiation frequency $\om$, the dechanneling 
length $\Ld(C)$  and Lindhard's critical angle $\Theta_{\rm L}(C)$ in the bent channel:
$\xi = \om\Ld(C)\Theta_{\rm L}^2(C)/2c$  (see~\cite{KKSG_demodulation} for the details).
The function $\alpha(\xi)$ is related to the 
real and imaginary parts of the first root (with respect to $\nu$) of the equation
\cite{KKSG_demodulation}
\begin{eqnarray}
F(-\nu,1,z)\Bigl|_{z=(1+\i)j_{0,1}\sqrt{\xi}/2} = 0
\label{Demodulation:eq.02}
\end{eqnarray}
where $F(.)$ stands for Kummer's confluent hypergeometric function (see, e.g., \cite{Abramowitz}). 

\begin{figure} [h]
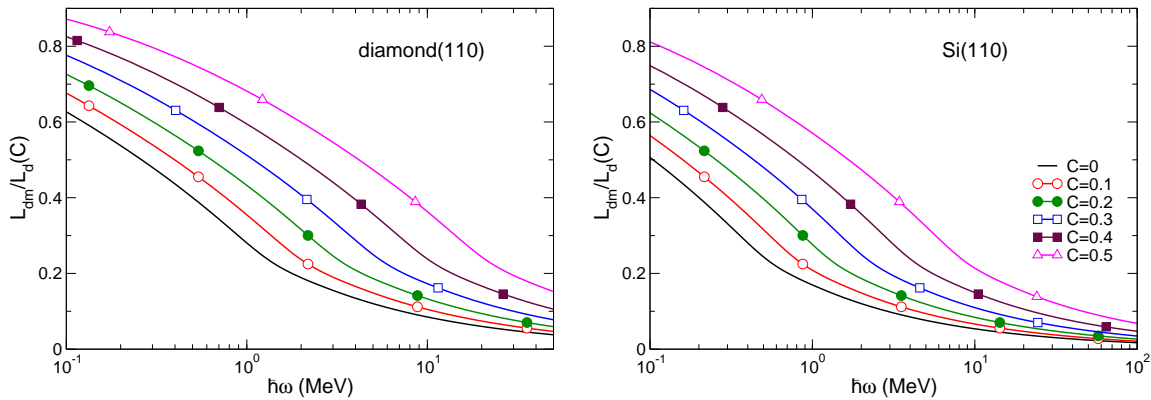

\centering
\includegraphics[scale=0.3,clip]{Figure10a.eps}%{Ldm_to_Ld-C110_v02.eps}
\hspace*{0.5cm}%
\includegraphics[scale=0.3,clip]{Figure10b.eps}%{Ldm_to_Ld-Si110_v02.eps}
\caption{
The ratio $\Ldm/\Ld(C)$ vs. photon energy for diamond(110) (left panel)
and silicon(110) (right panel) channels calculated for various values of bending parameter
$C$.
}
\label{Figure10.fig}%{Ldm_to_Ld-vs-om.fig}
\end{figure}

Eqs. (\ref{Demodulation:eq.01}) and (\ref{Demodulation:eq.02}) can be analyzed numerically to 
derive the dependence of the demodulation length on the radiation energy $\hbar\om$ for a 
particular crystal channel. 
The result of such analysis is illustrated by Fig. \ref{Figure10.fig} where the dependences of 
the ratio $\Ldm/\Ld(C)$ on the photon energy are presented for the (110) planar channels in diamond 
and silicon and for several values of the bending parameter $C$ as indicated.
To be noted, is that for all values of the bending parameter $C$ and over broad energy range of the 
emitted radiation, the demodulation length is noticeably less than the dechanneling one.

%%%%%%%%%%%%%%%%% 

To preserve the beam modulation during its channeling in a crystal and, 
as a result, to maintain the coherence of the radiation the crystal length $L$
must be less than the demodulation length:
\begin{eqnarray}
L \lesssim \Ldm < \Ld(C)\,.
\label{Demodulation:eq.03}
\end{eqnarray}
It follows from (\ref{Demodulation:eq.01}) that in a 
%PBC
periodically bent crystal
$\Ldm$ depends on the crystal type, 
on the parameters of the channel (its width, strength of the interplanar field),
on the bending amplitude and period, on the projectile energy and its type 
(these are "hidden" in $\Ld(C)$, $C$, and $\xi$) as well as on
the emitted photon energy (enters the parameter $\xi$).
Therefore, Eq. (\ref{Demodulation:eq.03}) imposes addition restriction on the CU length
as compared to the case of the CUR emission by the unmodulated beam.

In Ref. \cite{KKSG_demodulation} it is also shown that the phase velocity of the modulated beam along 
the CU channel is modified as compared to the unmodulated one. 
The modification changes the resonance condition which links the parameters of the undulator
and the radiated wavelength (energy). 
The expression for the fundamental harmonic frequency $\om\equiv\om_1$ acquires the following form
(compare with Eq. (\ref{Optimal_Length_CU:eq.02})):
\begin{eqnarray}
\om
= 
{ 2 \gamma^2 \over 1 + K^2/2 + \Delta_{\beta}^2/2}{2\pi c\over \lamu}
\label{Demodulation:eq.04}
\end{eqnarray}
where the additional term in the denominator is given by
\begin{eqnarray}
\Delta_{\beta}^2  
=  
4\gamma^2\Theta_L^2(C) \left(\beta(\xi) + {1 \over 2 j_{0,1} \sqrt{\xi}} \right)
\label{Demodulation:eq.05}
\end{eqnarray}
with $\beta(\xi)$ being another function related to the real and imaginary parts of the first root 
of Eq. (\ref{Demodulation:eq.02}) (details can be found in Refs. \cite{KKSG_demodulation,ChannelingBook2014}).
The quantity $\xi = \om\Ld(C)\Theta_{\rm L}^2(C)/2c$ depends on $\om$.
Therefore, Eq. (\ref{Demodulation:eq.04}) represents a transcendent equation which 
 relates $\om$ to the projectile energy and to the bending amplitude and period.

Analysis of the formulae written above shows that for given values of $\om$ and $\E$ all other quantities 
which characterize the CU and the demodulation process can be expressed in terms of a single independent 
variable, for example, the bending amplitude $a$.
Then, scanning over the $a$ values it is possible to determine the whole set of the 
parameters (these include $a$, $\lamu$, $C$, $\Ldm(C)$) which maximize the 
peak brilliance of the superradiant emission (see section \ref{SuperradiantCU}).

%%%%%%%%%%%%%%%%%% 
\begin{figure} [h]
\centering
\includegraphics[scale=0.45,clip]{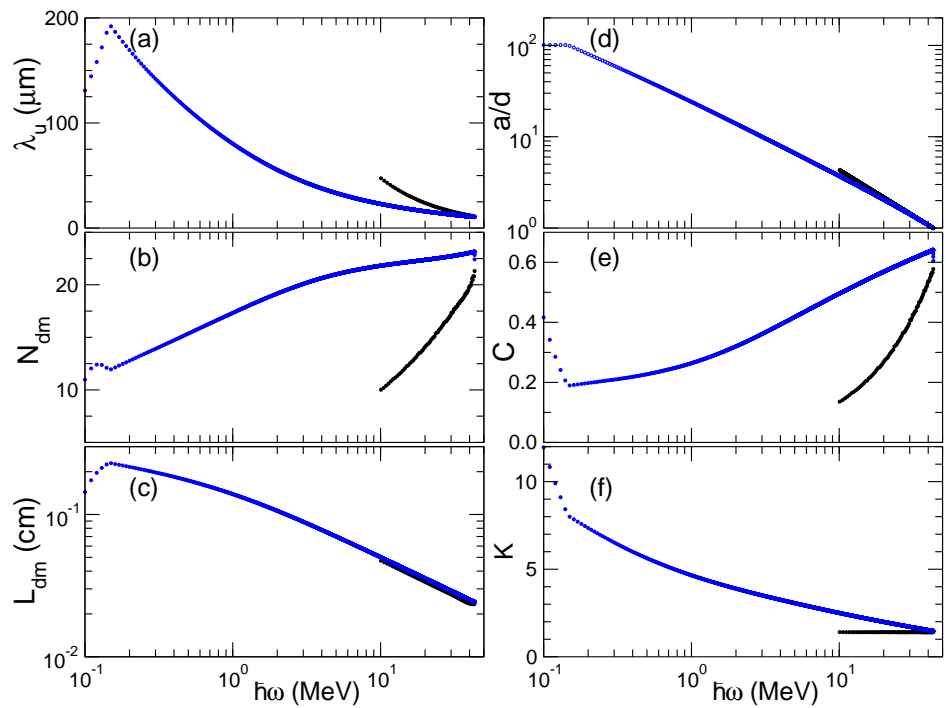}%{C110_FACET_v06b-2019.eps}
\caption{
Parameters of the diamond(110)-based CU, - $\lamu$, $a$ (measured in units of
the interplanar distance $d=1.26$ \AA), $C$, the undulator parameter $K=2\pi \gamma a/\lamu$,
the demodulation length $\Ldm(C)\equiv\Ldm$ and the number of periods within $\Ldm$,
$N_{\rm dm}=\Ldm/\lamu$ 
that ensure the highest peak brilliance of the radiation emitted by 
the fully modulated FACET-II positron beam.}
\label{Figure11.fig}%{C110_FACET_v05-2019.fig}
\end{figure}
% %%%%%%%%%%%%%%%%%%%%%%%%%%%%%%%%%%%%%

Figure \ref{Figure11.fig} shows the results of calculations of the parameters of the 
diamond(110)-based CU which maximize the peak brilliance of the radiation of energy $\hbar\om$ 
emitted by the FACET-II positron beam, see Table  \ref{ep-beams-2018.Table}.
The dependences presented correspond to the emission in the fundamental harmonic.
The crystal thickness was set to the demodulation length $L=\Ldm(C)$, graph (e).
The quantity $N_{\rm dm}$ stands for the number of undulator periods 
within the demodulation length, $N_{\rm dm}=\Ldm(C)/\lamu$.
Only the data corresponding to $N_{\rm dm}\geq 10$ are shown in the panels.
The dependences presented refer to the Large-Amplitude regime of the periodic bending,
which implies that the amplitude $a$ exceeds the interplanar distance $d$.

Noticing that the factor $2\pi/\lamu$ can be written in terms of the undulator parameter 
$K=2\pi\gamma a/\lamu$, one writes Eq. (\ref{Demodulation:eq.04}) as a quadratic equation with respect to $K$.
Resolving it one finds that $K$ is a two-valued function of $\om$, which is reflected in graph (f). 
As a result, all dependences presented contain two branches related to the smaller 
(black curves) and larger (blue curves) allowed values of $K$.

%%%%%%%%%%%%%%%%%%%%%%%%%%%%%%%%%%%%%%%%%%%%%%%%%%%%
\section{Brilliance of the Superradiant Emission in CU} \label{SuperradiantCU}

Powerful superradiant emission by ultra-relativistic particles channeled can be achieved if the probability 
density of the particles in the beam is (uniformly) modulated in the longitudinal direction with the 
period equal to integer multiple to the wavelength $\lambda$ of the emitted radiation 
\cite{Gover-EtAl_RPM_v91_035003_2019}.

To prevent the demodulation of the beam as it propagates through the crystal,
the crystal length $L$ must satisfy condition (\ref{Demodulation:eq.03}).
In a wide range of photon energies, starting with $\hbar\om\sim 10^2$ keV,
the demodulation length is noticeably less than the dechanneling length $\Ld$.
In addition to this, in this energy range the photon attenuation length $\La$ in 
silicon and diamond greatly exceeds the dechanneling length of positrons with energies 
up to several tens of GeV \cite{ChannelingBook2014}.
Therefore, on the spatial scale of $\Ldm$ the dechanneling and the photon attenuation
effects can be disregarded.

In what follows, we carry out quantitative estimates of the characteristics of the 
superradiant CU radiation (CUR) emitted by a fully modulated positron beam 
channeled in periodically bent diamond and silicon (110) oriented crystals in 
the absence of the dechanneling and the photon attenuation.
The beam represents a train of bunches each of the length $\Lb$ containing $\calN$ particles.
The crystal length (along the beam direction) is set to the demodulation length, $L=\Ldm$. 
The transverse sizes of a crystal are assumed to be larger than
those of the beam, i.e. than  $\sigma_{x,y}$.

For the sake of clarity, below we consider the emission in the first harmonics of CUR, 
see Eq. (\ref{Demodulation:eq.04})

Final width $\Delta \om$ of the CUR peak introduces a time interval $\tau_{\rm coh} = {1 / \Delta \om}$ 
within which two particles separated in space can emit coherent waves. 
Hence, one can introduce a coherence length \cite{SchmueserBook}
\begin{eqnarray}
L_{{\rm coh}} = c\tau_{\rm coh} 
= 
{\lambda \over 2\pi}{\om \over \Delta \om}
\label{Lcoherence:eq.02}
\end{eqnarray}
where $\lambda$ is the radiation wavelength, and the band-width (BW) $\Delta \om/ \om \approx 1/\Ndm$ 
with $\Ndm=\Ldm/\lamu$ standing for the number of periods within the demodulation length.

The number of the particles from the bunch which emit coherently is calculated as
\begin{eqnarray}
\calN_{\rm coh}
=
{L_{{\rm coh}} \over \Lb}\,\calN\,.
\label{Lcoherence:eq.03}
\end{eqnarray}
Their radiated energy is proportional to $\calN_{\coh}^2$.
The number of such sub-bunches is $\Lb /L_{{\coh}}$, therefore, the energy emitted by the whole 
bunch contains the factor $(\Lb /L_{\coh})\calN_{\coh}^2 = \calN\calN_{\coh}$.

Another important quantity to be estimated is the solid angle $\Delta\Om_{\coh}$
within which the waves emitted by the particles of the sub-bunch are coherent.  
This angle can be chosen as the minimum value from the natural emission cone of the first harmonics
$\Delta \Om = 2\pi \lamu/\Ldm$ and the angle $\Delta \Om_{\perp}$ which ensures transverse coherence 
of the emission due to the finite sizes  $\sigma_{x,y}$ of the bunch. 
Assuming the elliptic form for the bunch cross section one derives
$\Delta \Om_{\perp} \leq \lambda^2/4\pi\sigma_x\sigma_y$.
Therefore, the solid angle $\Delta\Om_{\coh}$ is found from
\begin{eqnarray}
\Delta \Om_{\coh}
=
\min\Bigl[\Delta \Om_{\perp},\Delta \Om \Bigr]\,.
\label{Coherence:eq.04}
\end{eqnarray}

The \textit{number of photons} $\Delta N_{\om}$ emitted by the bunch particles
one obtains multiplying the spectral-angular distribution of the energy emitted 
by a single particle by the factor 
$\calN\calN_{\coh}\,\Delta \Om_{\coh}(\Delta \om/ \om)$.
The result reads:
\begin{eqnarray}
\Delta N_{\om}
=
4\pi \alpha \,\calN \calN_{\coh}\,
\zeta
\left[J_0(\zeta)-J_1(\zeta)\right]^2
\Ndm \,
{\Delta\Om_{\coh} \over \Delta \Om}\,
{\Delta \om\over \om}\, .
\label{N_Photons:eq.04}
\end{eqnarray}
where $\zeta = (K^2+\Delta_{\beta}^2)/2(2+K^2+\Delta_{\beta}^2)$ with 
$\Delta_{\beta}^2$ defined in (\ref{Demodulation:eq.05}).

The number of photons emitted by the particles of the unmodulated beam in a CU of the same 
length and number of periods one calculates from  
Eq. (\ref{Optimal_Length_CU:eq.01}) written for $n=1$ by setting $\Neff=\Ndm$, 
substituting $K^2 \to K^2+\Delta_{\beta}^2$ and multiplying the right-hand side by $\calN$.
Comparing the result with Eq. (\ref{N_Photons:eq.04}) one notices that the enhancement factor
due to the coherence effect is $\calN_{\coh}\,{\Delta\Om_{\coh}/ \Delta \Om}$.

Another quantity of interest is the \textit{flux} $F_{\om}$ of photons. 
Measured in the units of $\Bigl(\mbox{photons}/\mbox{s}/ 0.1\% \mbox{BW}\Bigr)$,
it is related to $\Delta N_{\om}$ as follows:
\begin{eqnarray}
F_{\om} = {1\over 10^{3}(\Delta\om/\om)}\, {\Delta N_{\om} \over \Delta t_{\rm b}}
\label{N_Photons:eq.05}
\end{eqnarray}
where $\Delta t_{\rm b} = \Lb/c = e\calN/I_{\max}$ is the time flight of the bunch
and $I_{\max}$ stands for the peak current.

The \textit{peak brilliance}, $B_{\rm peak}$, of the superradiant CUR one obtains 
substituting $\Delta N_{\om}$ from (\ref{N_Photons:eq.04}) into Eq. (1) in the main 
text and using there peak current $I_{\max}$ instead of $I$.

%%%%%%%%%%%%%%%%%% 
\begin{figure} [h]
\centering
\includegraphics[scale=0.45,clip]{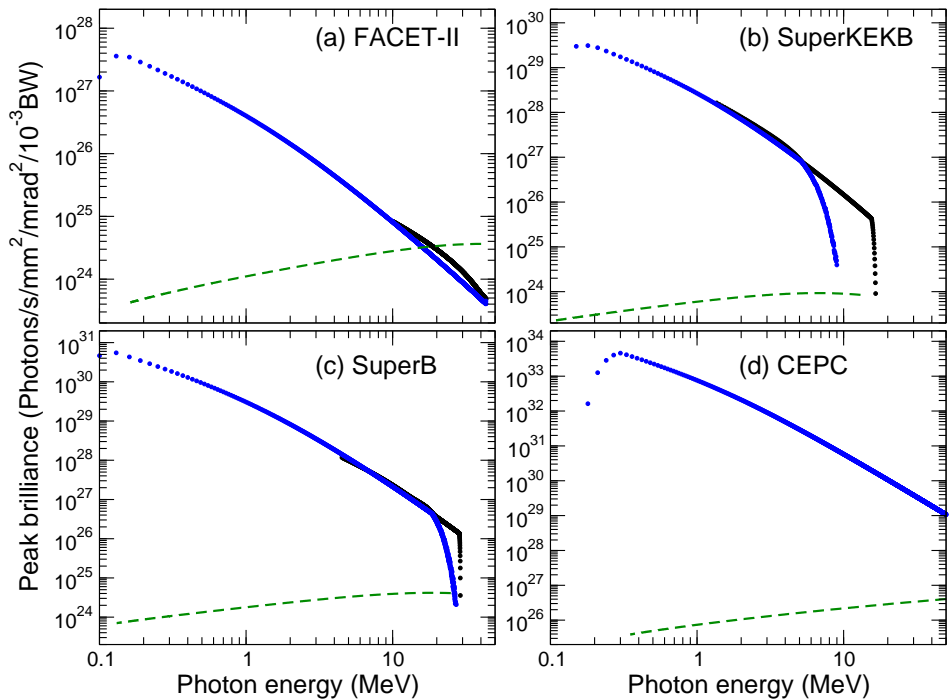}%{C110-CUR-CUL-all_v03.eps}
\caption{
Peak brilliance of superradiant CUR (thick curves) and spontaneous CUR (thin dashed curves)
emitted in periodically bent oriented diamond(110) crystal.
The graphs refer to four positron beams (as indicated).}
\label{Figure12.fig}%{C110-CUR-CUL-all.fig}
\end{figure}
% %%%%%%%%%%%%%%%%%%%%%%%%%%%%%%%%%%%%%
 
Figure \ref{Figure12.fig} shows peak brilliance of radiation formed in
the diamond(110)-based CU as functions of the first harmonic energy. 
Four graphs correspond to the positron beams (as indicated) the parameters of 
which are listed Table \ref{ep-beams-2018.Table}.
In each graph, the dashed line refers to the the emission of the spontaneous CUR 
formed in the undulator with optimal parameters, see Fig. \ref{Figure09.fig}. 
The thick curves present the peak brilliance of the superradiant CUR maximized by the proper
choice of the bending amplitude and period (as described in Section \ref{Demodulation}).
Two branches of this dependence, seen in graphs (a)-(c), are due to the two-valued 
character of the dependence of undulator parameter $K$ on the radiation frequency $\om$.
For the CEPC beam, graph (d), this peculiarity manifests itself in the frequency domain
beyond 40 MeV, therefore it is not seen in the graph. 

%%%%%%%%%%%%%%%%%%%%%%%%%%%%%%%%%%%%%%
\section*{References}
%%%%%%%%%%%%%%%%%%%%%%%%%%%%%%%%%%%%%%

%%%%%%%%%%%%%%%%%%%%%%%%%%%%%%%%%%%%%%%%%%
\end{document}